\documentclass{aa}      
\usepackage{graphicx}

\newcommand{\ma}{\mathrm}
\newcommand{\beq}{\begin{equation}}
\newcommand{\eneq}{\end{equation}}
\newcommand{\beqar}{\begin{eqnarray}}
\newcommand{\eneqar}{\end{eqnarray}}
\newcommand{\barn}{\begin{eqnarray*}}
\newcommand{\earn}{\end{eqnarray*}}

\def\simgt{\lower.5ex\hbox{$\; \buildrel > \over \sim \;$}}
\def\simlt{\lower.5ex\hbox{$\; \buildrel < \over \sim \;$}}

\begin{document}

\title{Stellar Envelopes as Sources of Broad Line Region Emission: New Possibilities
Allowed}

\author{G. Torricelli-Ciamponi\inst{1} \and P. Pietrini\inst{2}}

\institute{
Osservatorio Astrofisico di Arcetri, Largo E. Fermi 5, I-50125 Firenze, Italy 
\and
Dipartimento di Astronomia e Scienza dello Spazio, Universit\`a di Firenze,  Largo E. Fermi 2, 
I-50125 Firenze, Italy}

\offprints{P. Pietrini}
\date{Received / Accepted }

\titlerunning{Star Envelopes as Sources of BLR Emission.}
\authorrunning{G. Torricelli \and P. Pietrini}

\abstract{ In Active Galactic Nuclei (AGNs)
the presence of a star cluster around the central black
hole can have several effects on the dynamics and the emission
of the global system.  
In this paper we analyze the interaction of stellar atmospheres with
a wind outflowing from the central region of the AGN nucleus. Even a small
mass loss from stars, as well as possible star collisions, can give 
a non-negligible contribution
in feeding matter into the AGN nuclear wind. Moreover, stellar mass loss can 
produce envelopes surrounding stars that turn out to be suitable for reproducing 
the observed emission from the 
Broad Line Region (BLR). In this framework, the envelope can be  
confined by the bow shock arising from the interaction between  the  
expanding  stellar atmosphere and the AGN nuclear wind.
}

\maketitle

\section{Introduction}

It is generally accepted that the broad emission lines present in AGN spectra originate
in numerous, small and relatively cold gas concentrations in the 
neighborhood of the central black hole, the so-called Broad Line Region (BLR).
The  observed line properties directly indicate the existence 
of such gas concentrations; however, the physical nature 
of these structures is still a matter of debate. Up to the present
day, three possible classes of models have 
been envisaged: gas clouds, accretion disks, 
 and stellar wind envelopes. Each scenario suffers from still
unclarified problems and we refer to the recent review by Korista (\cite{korista})
for a presentation of the problem.

In general,  broad emission line regions in AGNs turn out 
to be characterized by a
relatively narrow range in ionization parameter
$U$, {\it i.e.},  
$U\equiv L/[4\pi c r^2n_*<h\nu>]
\simeq 0.01-1$ (Alexander \& Netzer \cite{alexander94},  
 Peterson \cite{peterson97}, Netzer \cite{netzer90}).
 In this expression $r$ is the distance from the central black hole, $n_*$ is 
 a representative value of the number 
density of the line emitting gas, $L$ is the 
ionizing radiation field luminosity from a central source, illuminating the line 
emitting region, and $<h\nu>$ is the average photon energy of 
the ionizing radiation field.
A first estimate of the conditions of the line emitting 
gas, such as $n_*$ and $U$, can be obtained from an analysis of 
the emission spectrum;
coupling this information with 
the evaluation of the luminosity and of the ionizing spectrum,
the relation above, defining $U$, can be inverted to give an estimate of the 
characteristic size of the BLR, or, better, the characteristic distance of the 
line emitting gas from the central continuum source, $r_{\ma {BLR}}$ (see Wandel et 
al. \cite{wandel99}). In a first, simplistic use of photoionization arguments,
under the assumption that 
the ionizing continuum shape is quite similar and 
both $n_*$ and $U$ are substantially the same 
for all the AGNs (see Kaspi et al. \cite{kaspi00}), 
a relationship, between the characteristic
BLR size and the AGN luminosity, of the type 
$r_{\ma {BLR}}\propto L^{0.5}$ would then be expected.
Indeed, this would just be  an order of magnitude evaluation, since there
are reliable indications that the BLR material density and the ionization parameter
are not constant at all even across the BLR extension of a single AGN (see, e.g., Kaspi et al. 
\cite{kaspi00}).
Nonetheless, this was actually what reverberation mapping studies of several AGNs seemed 
to suggest until very recently, implying
$r_{\rm {BLR}} \simeq 0.1~ L_{46}^{1/2}$ pc
(Netzer \& Peterson \cite{netzer}, Kaspi \cite{kaspi97})
(where $L_{46}$ is the AGN luminosity in units of $10^{46}$ erg/sec). 
In a reverberation technique study  of a sample of 17 quasars, combining their data 
with those available for Seyfert 1 galaxies, Kaspi et al. (\cite{kaspi00}) propose 
a significantly different relationship between $r_{\ma {BLR}}$ and the 
luminosity of the AGN, namely 
$r_{\ma {BLR}}\propto L^{0.7}$.  Although the exponent does not differ much from the 
previous one, it is remarkable that the ensemble of the data fit by the latter  relation, 
are not fit by the one with an 0.5 exponent. 

In a previous paper 
(Pietrini \& Torricelli \cite {pietrini}, hereafter Paper I), motivated by recent observations
supporting  radial outflows even in radio-quiet AGN's, such as Seyfert 1s, 
(Crenshaw et al. \cite{crenshaw99b},
Weymann et al. \cite{weymann97}), we have analysed the physical 
structure and characteristics of a global AGN outflow, presumably originating in
the very central regions and expanding out to large distances as a kind
of background/connection for various  observational components of the AGN
structure [ BLR, UV absorbers, X-ray "warm absorbers"].
A solution for the wind equations, accounting for the physical requirements
typical of central regions of Seyfert-like AGN, turned out to be 
possible only under rather specific conditions. 
One of the resulting constraints 
is that the outflowing wind may exist only if its density is  rather low.
The only possibility to increase the wind density is to
feed matter into the wind itself in a certain range of distances from the nucleus
(see Sect.~8.4 of Paper I for details).

As far as the BLR cloud model is concerned,  the presence of a nuclear wind  generally 
augments problems of cloud survival, rather than solving them, 
 unless the clouds are somehow comoving with the outflow
itself. In fact, one problem for the cloud
existence is the drag force at work between the moving clouds and a non-comoving confining 
medium. This drag force would rapidly disrupt the clouds. 
The observationally inferred information  on the cloud  
kinematics indicates that this motion is not radially directed and  
it must be characterized by a more complex velocity field. Also, the possibility of a Keplerian 
pattern has been suggested by Wandel et al. (\cite{wandel99}).
Since, for the reasons explained above, our choice is to work within the
scenario of a radially
directed nuclear outflow, the cloud survival problem would inevitably be worsened by the
presence of a radial nuclear wind.

For the stellar wind envelopes, since they are continuously fed by the stellar wind,  disruption is not
a problem; another velocity component
in their relative motion with respect to the interstellar medium does not constitute a problem either. 
On the contrary, we will show in this paper that  this new component in the relative
motion of the envelopes with respect to the surrounding medium can be a fundamental factor; 
in fact, it can contribute significantly to confine the plasma envelopes via
the formation of bow-shock fronts.

The above considerations, together with the necessity of feeding matter in the
AGN wind, induce us to invoke stellar envelopes
as  a possible interpretation of the nature of BLR
emitting plasma. In addition, the virial assumption of Keplerian
motions used to interpret the emission line width (Wandel et al. \cite{wandel99}) 
is in good agreement  with the stellar origin of BLR emission. Indeed, this possibility has
been widely discussed in the literature.  In particular,
after two pioneering works of Scoville \& Norman (\cite {scoville}) and Kazanas
(\cite{kazanas}), the star model has been reanalyzed by Alexander \& Netzer
(\cite{alexander94}, hereafter AN1),
and Alexander \& Netzer (\cite{alexander97}, hereafter AN2), 
who introduce ``bloated'' stars as stars characterized by a very extended envelope, 
but, due to the specific AGN environment, different from supergiant stars as 
known in the solar neighbourhood.
We refer to AN1 for a wider presentation of the scenario and its problems.

In their papers, Alexander \& Netzer (AN1,AN2) present different possibilities
for confining ``bloated'' star 
envelopes and conclude that
the more effective one is tidal disruption by the black hole.  However, AN1 and AN2 
assume in their works that the stellar winds expand into vacuum, asserting that this 
does not constitute a limitation on the validity of the analysis.
On the contrary, in this paper we analyze whether and how the presence of a nuclear 
AGN wind, such as the  
one described in   Paper~I, can have an influence on the BLR 
physical structure. In particular,
by inserting mass losing stars in their specific environment and 
investigating the interaction of the stellar wind with the AGN wind,
we introduce another confining mechanism for the star envelopes. We then compare
it to  both the tidal one  and to the other possible confinement mechanisms investigated by 
AN1 and AN2.  Finally  we analyze the consequences for BLR parameter values. 

In Sect.~2, we discuss confinement of the expanding stellar envelopes and 
introduce the bow-shock formation mechanism as a further means of defining an outer 
boundary for these envelopes.
Section~3 is devoted to the connection and interaction between the AGN nuclear wind, 
as developed in Paper~I, and the central star cluster, with colliding and mass losing stars, 
in terms of the resulting mass deposition into the AGN nuclear wind. 
In nSect.~4, we describe the physical parameters  that turn out to be involved 
in the computation of the model, their significance in the outcome 
determination and the general 
properties of a stellar envelope necessary to 
produce the observed line emission.  We then identify a sort of ``observational test''
for the results of a BLR model, through the comparison with typical values inferred from 
observations of global BLR parameters, such as the covering factor, the 
characteristic BLR radius, 
the ionization parameter, and the fact that no broad forbidden lines are 
observed. Section~5 discusses  how the identified general 
requirements put physical limitations on our model parameters.
The general results of our model are presented in Sects.~5.1, 5.2, 5.3 
for three distinct 
cases,  that differ basically in the the choice of the mass loss rates for the various types of 
stars in the central cluster.
Section 6 is devoted to  general discussion of the main results, 
including a comparison with the works of other authors applying different models to infer 
the general properties of  
the BLR of  a specific source, NGC~5548, which is well representative of the Seyfert~1
class of AGNs.
Finally, in Sect.~7 we summarize our results and the general features that 
characterize our interpretation of the BLR, 
highlighting the qualities and the limitations of the model.

\section{Confinement of the stellar envelopes:
Bow shocks around moving stars versus other mechanisms}

When a mass-losing star moves rapidly through the interstellar 
medium, its wind interacts with the surrounding plasma, 
giving rise to an elongated bow shock. 
The physics of these shock fronts has been widely
studied in different contexts (see e.g Perry \& Dyson \cite{perry},
Van Buren et al. \cite{buren} and references therein) for different purposes.
In general, we expect that the expanding stellar wind 
can develop a shock front, in which the wind kinetic energy is
partially converted and radiated away. Therefore, in a stationary
configuration, the shock position is determined by the
equilibrium of the  external pressure  and that of the stellar wind.
The resulting shock geometry is a cometary-like tail
elongated in the direction opposite to that of the star motion.

In our case the situation is more complex since 
stars move around the AGN nucleus with a velocity ${\bf V}(r)$
 and  are embedded in a radial nuclear wind  characterized by a number density $n_{\rm w}(r)$
and expanding with velocity ${\bf v_{\rm w}}(r)= v_{\ma w}(r) {\bf \hat e_{\ma r}}$.
The AGN wind velocity $v_{\rm w}(r)$,  number density $n_{\ma w}(r)$ and temperature
$T_{\ma w}(r)$ profiles are those computed through the resolution of the system of stationary
hydrodynamical equations,  described in Paper~I, including the appropriate mass deposition.
The nuclear wind velocity $v_{\ma w}$  undergoes the transition from sub-sonic to 
super-sonic at 
a distance $r_{\rm c}$
from the AGN nucleus, where $r_{\rm c}/r_{\ma g} \ge 1/4 (c/c_{\rm s})^2$,
with $c_{\rm s}$ being the local sound speed,  $c$ the light speed, and 
$r_{\ma g}\equiv 2GM_{\ma {BH}}/c^2$ 
the black hole Schwarschild radius.

For the stellar velocity we adopt a Keplerian motion (consistent with
 recent interpretations of emission line width (Wandel et al. \cite{wandel99})):
\beq 
V(r)=\sqrt {GM(r)\over\ r},
\label{velstar}
\eneq
where $M(r)$ is the total mass (central black hole plus star cluster) within the radius
$r$, assuming its distribution is spherically symmetric.
The velocity  of the stars is generally supersonic inside the 
nuclear wind sonic radius, while, of course, 
$v_{\rm w}$ is supersonic outside.
Hence, in our case the bow shock will have a different geometry
depending on the regime in which it is computed:
cometary-like shocks with the tail in the direction
opposite to the star motion (when the ram pressure contribution
is induced by the star motion, $m_{\rm H}n_{\rm w}V^2$);  or
with the tail extending radially outwards from  the central black hole
(when  the ram pressure contribution is due to the AGN wind:
 $m_{\rm H} n_{\rm w} v_{\rm w}^2$); or a combination of the two.
It is beyond the scope of this paper to investigate
the detailed physics  and geometry of this configuration; instead, what  is important in this
context is the existence of a region around the star where the  stellar 
wind can be confined by this mechanism.
For the sake of simplicity we have chosen to represent this complex wind envelope region
schematically as a quasi-spherical structure, so that its geometrical extension can be characterized 
simply by a radius defining the radial distance of the envelope boundary from the star.

In order to derive this envelope radius, first we have to parameterize the 
physical properties of the expanding envelope, namely  give  a prescription for the 
density of the gas in the envelope itself and for its expansion velocity. 
Our simplifying choice is to adopt a radial power law behaviour for the 
stellar atmosphere expansion velocity:
\begin{equation}
  v_*(R)=v_0 \left(R\over R_*\right)^{\alpha}~~~~~~~{\rm for}~~R>R_*,
\label{velocR}
\end{equation}
where $R$, the distance from the star center,
 and $R_*$, the radius of the star, are both expressed in the same length
units; this relation is taken as independent of the distance, $r$, 
of the star from the central black hole.
Taking into account continuity equation for the envelope gas, we can derive
its density as a function of $R$ as well:
\beq
n_{\ma {env}}(R) = {\dot M_*\over 4\pi~m_{\rm H} R^2 v_*(R)} =
             \left({\dot M_*\over 4\pi~m_{\rm H}v_0 R_*^2}\right)
			 \left({R \over R_*}\right)^{-(\alpha+2)}.
 \label{nenv}
\eneq

Once these quantities are defined,
the order of magnitude of the distance from the star,
$R_{\ma {bow}}$, at which the bow shock can be formed, 
can be determined by using the  standard methods, that is from the two equations
describing the stellar mass loss rate (i.e., continuity equation again) and the balance between
the total pressure  (thermal+ram pressure) of the stellar wind and
that of the ambient medium,  all evaluated at $R_{\ma {bow}}$:
\begin{equation}
{\dot M_*}=4\pi~m_{\rm H} ~R_{\rm bow}^2 n_{\ma {env}}(R_{\rm bow})v_*(R_{\rm bow})
 \label{massl}
\end{equation}
\begin{equation}
P_*+m_{\rm H} n_{\ma{env}}(R_{\rm bow})v_*^2(R_{\rm bow})=P_{\rm w}+P_{\rm ram}. \label{ptoeq}
\end{equation}
Here we assume that $P_* = n_{\ma{env}}kT_*/\mu_*$ 
and $P_{\rm w} = n_{\rm w}kT_{\rm w}/\mu_{\ma w}$ are the thermal pressures
of the star and of the AGN winds, respectively, where $n_{\ma i}\equiv \rho_{\ma i}/m_{\ma H}$
(for ${\ma i} = {\ma{env}},{\ma w}$), with $\rho_{\ma i}$ the mass density. We have taken 
$\mu_*=\mu_{\ma w} =0.6$, as expected for a fully ionized plasma with cosmic abundances
(see Allen \cite{allen}); note that in the following we indicate both $\mu_*$ and 
$\mu_{\ma w}$ as $\mu$.
Also, $v_*$ is the 
expansion velocity of the stellar atmosphere
as given by Eq.~(\ref{velocR}) for $R=R_{\ma {bow}}$.
The ram pressure of the external medium
can be identified with the maximum of that due to the star motion and that
due to the AGN wind:
$$P_{\rm ram}=Max[m_{\rm H}n_{\rm w}V^2,m_{\rm H} n_{\rm w}v_{\rm w}^2]=m_{
\rm H} n_{\rm w} v_{\rm Max}^2.$$
The two equations above (Eq.~(\ref{massl}) and Eq.~(\ref{ptoeq}))
turn out to depend on the distance of the star (whose envelope we are describing)
from the central black hole, $r$, since they contain physical quantities depending on $r$ 
itself. These are the ones defining 
the AGN nuclear wind ($n_{\ma w}(r), T_{\ma w}(r),v_{\ma w}(r)$), and the Keplerian velocity of the 
star in the local gravitational field, as given by Eq.~(\ref{velstar}).
Also, these quantities can be combined so as to define 
global envelope properties as functions of $r$ as well. 
In fact, Eq.~(\ref{massl}) can be rewritten as
\begin{equation}
n_*(r)\equiv n_{\ma {env}}(R_{\ma{bow}}) = {{\dot M_*} \over 4\pi ~m_{\rm H} R^2_{\rm bow}v_* 
 }\label{next}
\end{equation}
and, making use of Eq. (\ref{next}),
  Eq.~(\ref{ptoeq}) becomes 
\begin{equation}
R_{\rm bow}^2(r)={{\dot M_*} \over 
4\pi ~m_{\rm H} n_{\rm w}(r)v_* } ~{ m_{\rm H} v_*^2+
kT_*/\mu  \over m_{\rm H} v_{\rm Max}^2(r)+kT_{\rm w}/\mu }
 ,\label{bow}
\end{equation}
where $v_*$ indicates the value of the stellar atmosphere velocity at its boundary $R_{\ma{bow}}$,
as in Eq.~(\ref{ptoeq}). Note that in Eq.~(\ref{next})
we have redefined $n_{\ma {env}}(R_{\ma {bow}})$,
the density of the stellar envelope at its external boundary, 
as $n_*(r)$, explicitly showing its dependence on $r$, since its value 
does depend on the distance of the envelope from the black 
hole, as mentioned above. 
In the following, unless otherwise specified, 
we use $n_*(r)$ to indicate	a reference number density value 
characterizing the stellar atmosphere.
As for the stellar atmosphere gas temperature, here we assume for simplicity that
it is constant, i.e. independent both of $r$ and of $R$; this assumption and its validity 
within the present framework are briefly discussed in Sect.~5.
The two relationships shown above define the density and the extension of the stellar  
wind envelope as  functions both of the stellar wind velocity and temperature,
and of the external confining plasma parameters. Since 
the stellar wind velocity depends on $R_{\rm bow}$, Eq.(\ref{bow}) is in general
an implicit equation.

As Alexander \& Netzer (AN1, AN2) have shown, several other mechanisms 
in principle can be competing for the definition of the physical extension 
of the stellar wind envelope. 
In the following we briefly outline them, in order to compare the characteristic envelope 
radius values defined by each mechanism at any given radial distance $r$ from the black hole 
with $R_{\ma {bow}}$, {\it i.e.} the distance from the star at which the bow shock, 
due to the interaction of the stellar wind with the nuclear AGN outflow, would form.
The various physical processes that are to be examined produce a stellar wind envelope extension
that is characterized by a different dependence on the radial distance $r$ from the 
central black hole.  Therefore, for each value of $r$, we shall finally 
identify the effective extension of the envelope itself with the smallest radius 
obtained, thus selecting the most efficient mechanism for stellar wind confinement.

A first effect we need to take into account is the possibility of tidal disruption 
of the outer layers of the stellar envelope, due to the black hole gravitational pull.
In this case the size of the stellar envelope is given
by (see AN1)
\beq
R_{\rm tidal} \simeq 2 \left({M_* \over M_{\rm BH}}\right)^{1/3} r,
\label{rtid}
\eneq
showing a simple linear dependence on the distance from the black hole.

Another possible definition of an upper limit to the stellar wind extension
can be determined by the reasonable assumption of the existence of a finite mass
in the stellar wind. Again following AN1, 
we suppose that
the total mass in the stellar wind must not exceed 0.2~M$_\odot$;
this leads to the condition
\beq
\int_{R_*}^{R_{\ma {mass}}} \dot M_* dt = \int_{R_*}^{R_{\ma {mass}}}{\dot M_* 
  \over v_*}dR = 0.2 M_\odot,
\label{rmass1}
\eneq
which, for a given stellar wind velocity field $v_*(R)$,  defines $R_{\ma{ mass}}$ as 
the maximum stellar wind  size allowed for, due to the limit on the stellar 
wind mass content. 
Assuming that Eq.(\ref{velocR}) defines the velocity profile $v_*(R)$ for the stellar wind, 
we can explicitly solve the integral to obtain 
\beq
R_{\ma {mass}} = \left [\left({(1-\alpha) 0.2M_{\odot}v_0\over \dot M_*} +
    R_*\right)R_*^{-\alpha}\right ]^{1/(1-\alpha)};
\label{rmass2}
\eneq
notice that in this specific case the stellar envelope extension would not depend 
on the radial distance $r$ from the central black hole.

Comptonization effects, due to the central ionizing continuum heating the outer 
layers of the stellar envelope, should also be taken into account in the determination 
of the external boundary of the envelope itself. This  process (see Kazanas \cite{kazanas},
and AN1) induces complete ionization of the outer layers of the stellar wind,
that become part of hot ``Comptonized'' phase. It is effective 
on the denser stellar wind down to layers in which the density of the envelope material reaches
a critical value, $n_{\ma {Comp}}$, above which, that is deeper into the envelope, a 
``cool phase'' is still allowed (Kazanas \cite{kazanas}; Krolik, McKee, \& Tarter 
\cite{krolik}), and therefore the stellar wind survives as such.
This density value is defined by a critical value, $\Xi_{\ma c}$,  of the 
``ionization'' parameter $\Xi\equiv L_{\ma{ion}}/(4\pi r^2c n k T)$ for the 
existence of a ``cool'' phase, for a given ionizing radiation field, and it is 
\beq
n_{\ma {Comp}} (r) = {1\over \Xi_{\ma c}} {L_{\ma {ion}}\over 4\pi c k T_*}{1\over r^2}.
\label{ncrit}
\eneq
This critical value of the density defines a boundary radius for the stellar wind, that we can call ``Comptonization''
radius, $R_{\ma {Comp}}$, by setting $n_{\ma{env}}(R)=n_{\ma {Comp}}$ in the continuity equation
(see Eq.(\ref{nenv})).
We thus obtain
\beq
R_{\ma {Comp}} = \left[{\Xi_{\ma c} \dot M_* c k T_* r^2\over m_{\ma H}v_0 
	R_*^{-\alpha}L_{\ma {ion}}}\right]^{{1\over 2+\alpha}}.
\label{rcomp}
\eneq

\noindent
In equations (11) and (12)  the quantities $T_*$ and $\Xi$ cannot be freely
chosen, since they depend on each other, their specific relationship, $T_*(\Xi)$,
deriving from the analysis of possible equilibrium configurations existing in
a photoionized gas (Krolik, McKee, \& Tarter \cite{krolik}).
Since $T_*(\Xi)$ depends on the assumed continuum radiation illuminating the gas, as well as 
on other heating mechanisms possibly present
and on cooling processes, its resulting detailed shape 
turns out to be different in literature depending on the specific assumptions of the authors. 
Krolik (\cite{krolik99}) shows a $T(\Xi)$ curve, obtained with the composite continuum spectrum 
from Elvis et al. (\cite{elvis94}), in which $\Xi_{\ma c}\simeq 10$ and the corresponding 
cool phase limit temperature is $\sim 5\times 10^4$~K.
In other studies (see Krolik \cite{krolik02}),
the ``cool'' phase ($T< 10^5$K)
 exists for $\Xi \leq 28$ for the continuum spectrum considered by 
Kaspi et al. (\cite{kaspi01}), whereas for Krolik \& Kriss (\cite{krolik01}) it is present 
only for $\Xi \leq 12$. 
The temperature corresponding to these upper limits  for the ionization parameter $\Xi$ 
is around $5\times 10^4$K for the 
two cases.
It is somehow difficult to make a consistent choice for the values of $\Xi_{\ma c}$ and $T_*$ to be used
in Eq.~(12) without computing directly the specific
equilibrium, therefore we choose to rely  on the assumptions of one of the papers in literature, 
namely  the one of Kazanas (\cite{kazanas}). In this work, the author computes the critical density 
for the transition to the Comptonized phase (see Eq.~(11)) taking  the critical value of 
the ionization parameter as $\Xi_{\ma c}=10$ and a temperature ($T_*$ in our notation)
equal to $3\times 10^4$~K (Kazanas \cite{kazanas}). Given the results of most recent studies and calculations
(some of which have been cited above), 
these values can be regarded as sort of ``lower limits'' for the possible range of critical 
ionization parameter and/or temperature. However, in the present context we maintain this choice 
as  representative of the critical condition, since, owing to the 
functional dependence of the limiting radius, $R_{\ma{Comp}}$, as expressed in Eq.~(12), 
``lower limit'' values for $\Xi_{\ma c}$ and the corresponding $T_*$ guarantee 
a ``lower limit'' value for the Comptonization radius as well. In fact, we are evaluating 
this radius to compare the efficiency of this confinement mechanism with  that of the other relevant ones.
If indeed Comptonization turned out to be the effective confinement mechanism for the stellar
envelopes contributing to the line emission in the BLR, a more accurate analysis of appropriate
parameters to be used in the precise evaluation of $R_{\ma {Comp}}$ would be required. 
Nevertheless, if, as it
will be proved to be the case in all the conditions of interest for envelopes contributing to the
construction of our BLR model,
the physical extension of the stellar envelopes is not determined by $R_{\ma {Comp}}$, the present
evaluation is fully representative. Also, it is important to note that, if this is the case, 
the Comptonization radius limitation itself is no longer 
physically relevant to the problem. 
In fact, outside the otherwise confined stellar envelope, the gas is mixed up with the surrounding 
nuclear wind and shares its physical properties.

Finally,  AN1 mention another limit for the external radius of the 
stellar wind envelope, deriving from the requirement that the stellar wind velocity, 
when it is  increasing outwards, 
does not reach values larger than $0.1c$. This is a rather artificial constraint, 
defining a radius $R_{\ma v}$, 
which, for our choice of the expansion velocity of the stellar atmosphere, 
is $R_{\ma v} = R_* \left({0.1c/ v_0}\right)^{1/\alpha}$, and
 should apply only in extreme cases. 

In summary, 
we  define the effective stellar envelope extension 
at each distance $r$ from the central black hole as  the smallest of
the various radii determined above:
\begin{equation}
R_{\ma{ext}}={\rm min} [R_{\ma{bow}}, R_{\ma{tidal}}, R_{\ma {mass}}, R_{\ma{Comp}},R_{\ma v}].
\label{ext}
\end{equation}
From our computations, for appropriate BLR models  
the most efficient confinement mechanism turns out  to be the bow-shock formation, thus 
selecting $R_{\ma{bow}}$ as the stellar envelope extension in the equation above.
In this case, for larger distances from the star center, i.e. for $R>R_{\ma{bow}}$, 
the gas physical properties are just those of the nuclear wind gas (i.e., $n=n_{\ma w}$ and 
$T=T_{\ma w}$), since the shock transition is a sharp one.
Using the notation of Eq.~(\ref{next}), the characteristic 
number density $n_*$ can be derived from the continuity 
equation [Eq.(\ref{next})
with $R_{\ma{ext}}$ the place of $R_{\ma{bow}}$] 
\begin{equation}
n_*(r)= {{\dot M_*} \over 4\pi ~m_{\rm H} R^2_{\rm ext}v_*}.
\label{cont}
\end{equation}

In the current framework,  
the stellar wind gas generates the line ~emission via  central radiation re-processing,
 hence the quantities  $R_{\rm ext}$ and $n_*$ 
determine   the observed properties of the emission lines.
An AGN wind which  generates  suitable values of the quantities $n_*$ and $R_{\rm ext}$
through Eqs.  (\ref{cont})  and (\ref{ext})  can then  represent the requested link between the
central engine and 
other structural components of the AGN itself, such as the BLR.

\medskip

\section{Mass deposition in the AGN wind}

In a picture where the AGN wind expands among mass-losing stars,
the presence of the stars will influence the AGN wind dynamics
by adding mass to the wind itself. Two different sources of mass deposition
are possible: the mass directly lost by the stars and that 
stripped by star collisions.
This  implies that nuclear wind models in such a
framework must allow for  mass deposition increasing the total mass flux
along the radial coordinate. 
Indeed, this possibility has been taken into account in Paper I,
in order to maintain the mass density in the wind at non-negligible
levels in the outer region,  by accounting for a possible 
mass source along the wind. In this case the total mass flux
(mass/time, dimensionally) is not a constant, but  is rather a function 
of the radial coordinate $r$, namely
$\dot M_{\ma w}= 4\pi m_{\ma H}n_{\ma w} v_{\ma w} r^2= 4\pi A(r)= 4\pi A_0f_{\ma m}(r)$.
In this expression, $A(r)$ and the constant $A_0$
are dimensionally mass per unit time
per unit solid angle, whereas $f_{\rm m}(r)$ is
a dimensionless function of radial distance from the central black hole
(see Paper I, Sect.~3 for details).
The constant $A_0$ is defined as  
$A_0 \equiv n_{\ma c}m_{\ma H}r_{\ma c}^2 c_{\ma{sc}}$, 
where $c_{\ma{sc}}$ is the sonic speed at the critical point $r_{\ma c}$
and $n_c$ is a
characteristic value of the wind number density, which is related to the number density 
value at the critical point by the following relation: 
$n_{\ma w}(r_{\ma c}) = n_{\ma c}f_{\ma m}(r_{\ma c})$.
With the notation defined above, the mass source term [mass/time/volume/solid angle)] is 
$\nabla\cdot(m_{\ma H}n_{\ma w}{\bf v_{\ma w}})=(dA/dr)/r^2$. 

In order to evaluate the amount of mass fed into the wind
by stellar mass loss and by star collisions, we have computed the
contribution from each process as described  in the following.
The total contribution, up to a distance $r$ from the central black hole, 
to the mass flux in the nuclear wind from these 
external sources is 
\beq
\dot M_{\ma {inp}} = 4\pi[A(r)-A(r_0)],
\label{minp}
\eneq
where $r_0$ is the 
radius at the base of the wind, and, consequently, $4\pi A(r_0)$ is the 
``intrinsic'' mass flux in the nuclear wind, at its starting point.

Before proceeding, we need to define a schematic 
description of the spherically symmetric compact dense
stellar cluster that we suppose to be located at typical BLR distances
from the central black hole of the AGN. In the present work, we 
assume to represent it in a strongly simplified way, that is 
considering just three different stellar components for it,
main sequence stars (MSs), red giants (RGs), and supergiants (SGs).
These three stellar components lose mass to the external interstellar
medium, which is here the nuclear wind plasma, both due to their own 
stellar winds, whose relevance depends of course on the specific stellar 
component, and because of the possible collision-like encounters
that the stars undergo.  
We also define the relative percentage of RGs and SGs, with respect 
to the total stellar number density of the cluster, $\rho_*$, as
$f_{\ma {RG}}$ and $f_{\ma {SG}}$ respectively, so that
$(\rho_*)_{\ma {RG}} = f_{\ma {RG}}\rho_*$ and 
$(\rho_*)_{\ma {SG} }= f_{\ma {SG}}\rho_*$, 
whereas, for reasonably small values of the fractions 
$f_{\ma {RG}}$ and  $f_{\ma {SG}}$, we can take 
$(\rho_*)_{\ma {MS}}\simeq \rho_*$, {\it i.e.}, $f_{\ma {MS}}\simeq 1$.
In the computation of our models, we have usually assumed a ``standard''
value for the relative fractions of RG and SG stars, that is 
$f_{\ma {RG}}= 0.01$ and  $f_{\ma {SG}}= 10^{-4}$ (see AN1), but we 
keep those fractions generically as parameters of the problem in what follows.

As for the stellar mass loss from the stars of 
a  central cluster characterized by a stellar number density $\rho_*(r)$
and a composition as defined above, 
the total contribution to the AGN nuclear wind  mass loading rate, up to a 
radial distance $r$ can be expressed as 
\beq
\dot {M}_{*\ma{wind}}(r) \simeq ~~~~~~~~~~~~~~~~~~~~~~~~~~~~~~~~~~
\label{minpw}
\eneq
$$ 
\int_{r_0}^r\left \{(\dot M_*)_{\ma{MS}}f_{\ma{MS}} + (\dot M_*)_{\ma{RG}}f_{\ma {RG}}
 + (\dot M_*)_{\ma{SG}}f_{\ma {SG}}\right\} \rho_* 4\pi r^2  dr,
$$
 where $(\dot M_*)_{\ma i}$ represents the mass loss rate in the wind of the ``i''th
 stellar component of the cluster.
 For reasons that will be extensively discussed in the 
next sections and are related to the very specific and extreme environment in which 
the central stellar cluster is located (see, for example, Taylor \cite{taylor}),  
here we allow for a 
possible non-negligible contribution from main sequence (MS) stellar winds. 

As already mentioned, the other contribution to mass deposition into the AGN nuclear 
wind  comes from stellar collisions. 
Following AN1 and Murphy et al. (\cite{murphy}),
we only take into account 
collisions between stars of the same type, thus neglecting the contribution 
due to MS-RG and MS-SG star collisions, which are characterized by a much 
smaller fractional mass loss with respect to collisions between same type stars.
In the hypothesis that for each collision a fraction $f_{\ma c}$ (that we take to be 0.1 
following AN1 and Begelman \& Sikora \cite{begelman92}) of 
one solar mass
of matter is stripped from the colliding stars and fed into the wind,
the total mass deposition rate due to collisions occurred up to radial distance $r$,
  $\dot {M}_{*\ma{coll}}(r)$, is
$$\dot {M}_{*\ma{coll}}(r) \simeq ~~~~~~~~~~~~~~~~~~~~~~~~~~~~~~~~~~$$
\beq
0.1\left [(\tau_{\ma{coll}})_{\ma{MS}}(r) 
+ (\tau_{\ma{coll}})_{\ma{RG}}(r)+(\tau_{\ma{coll}})_{\ma{SG}}(r)\right] M_{\odot}/{\rm yr},
\label{minpc}
\eneq
where $(\tau_{\ma{coll}})_{\ma i}$ is the star collision rate (here expressed as 
number of collisions per year) 
for the stellar component ``i'' defined  as:
\begin{equation}
 (\tau_{\ma{coll}})_{\ma i}(r) = 4 \pi^2 (R_*)_{\ma i}^2  \int^{r}_{r_0} (f_{\ma i}\rho_*)^2
 r^2 V(r) dr~~
\label{tau}
\end{equation}
with $(R_*)_{\ma i}$  the stellar radius of stars of type ``i'', and $f_{\ma i}$ the fraction
of stars of type ``i'', that is assumed to be equal to unity for the case ``i''$=$ MS, that is for 
the main sequence star component.

For the definition given above of $\dot M_{\ma{inp}}(r)$, 
the following relationship must hold
\begin{equation}
\dot M_{\ma{inp}}(r)= {\dot M_{\ma{*wind}}(r)}+ {\dot M_{\ma{*coll}}(r)}.
\label{starden}
\end{equation}
Indeed, we are interested in the equation that can be obtained from the previous one by 
deriving it with respect to $r$ and dividing by $4\pi$, and that represents the equation for the sum of the 
mass sources from stellar matter expulsion 
processes at distance $r$,
giving the total mass source function per unit solid angle $(dA(r)/dr)/r^2= (A_0/r^2) df_{\ma m}/dr$ 
for  the nuclear wind (see Paper I).
This equation turns out to be the following:
\begin{equation}
A_0 {df_{\rm m}(r)\over dr}{1\over r^2}=F(r),
\label{fr}
\end{equation}
where the function $F(r)$ is defined as follows
$$F(r)= \left [(\dot M_*)_{\ma{MS}}f_{\ma {MS}} + (\dot M_*)_{\ma{RG}}f_{\ma {RG}}
 + (\dot M_*)_{\ma{SG}}f_{\ma {SG}}\right] \rho_*(r) +$$
 $$~~~~0.1 M_{\odot} \pi V(r)\rho_*^2(r)\times~~~~~~~~~~~~~~~~~~~~~~~~~~~~$$
\begin{equation}
~~~~~~~~~~ \left [(R_*^2)_{\ma {MS}}f_{\ma {MS}}^2 +
(R_*^2)_{\ma {RG}}f_{\ma {RG}}^2+
(R_*^2)_{\ma {SG}}f_{\ma {SG}}^2\right].
\label{starden2}
\end{equation}
From Eqs.(\ref{fr}) and (\ref{starden2}), given $\rho_*(r)$, it is evident that for a 
given choice of the physical parameters of each stellar component,
$A(r)$ can be derived
by solving the equation itself. 
As it will be discussed in a following Section, 
the prescription of $\rho_*(r)$ is constrained by several factors, but 
$\rho_*(r)$ must be chosen so as to obtain a mass source function with a sufficiently smooth
behaviour.  This is relevant for the  successful integration
of the nuclear wind equations. Of course, the resulting mass source function depends on the 
mass loss rates that we adopt for our model cluster stars, 
as well as on the chosen values of $(R_*)_{\ma i}$, and of the population fractions $f_{\ma i}$, 
for the three stellar components of the cluster;  
this is  also extensively discussed in the following Sections.

A numerical comparison 
of the two terms on the right hand side of Eq.(\ref{starden2})
can show whether one of the two processes of mass deposition  in the nuclear wind 
predominates and, in that case, which one it is, as a function of $r$.
For a given $\rho_*(r)$, the result of this analysis strongly depends on the choice of the mass loss rates
of the various types of stars in the model cluster. 

\medskip

\section{Model  parameters}
\medskip
\subsection{Free parameters in the model}

As it is apparent from the discussions in Sects. 2 and 3,
in order to derive the effective stellar envelope extension and
its density through Eq.s(\ref{cont}) and (\ref{ext}), the
wind model of Paper I must be extended to take into account
the mass deposition function (as derived from Eqs.~(\ref{fr}) and (\ref{starden2})) and
 to include the computation of $R_{\rm ext}(r)$ and $n_*(r)$.
This process implies the introduction of a number of parameters that can
be divided in three classes, each associated with a different
aspect of the problem. These parameters are summarized in the following
for the sake of clarity. As described below, some of these parameters have almost no
influence on the final results, some have been fixed to standard values and
some other represent the control keys to variations in the outcome 
of our model.

{\it a) Parameters related to the AGN nucleus and its wind:}
central radiation field luminosity, $L$,
characteristic number density, $n_{\ma c}$, and temperature, $T_{\ma c}$,
 of the wind plasma at the critical point (that is at the distance
at which the wind outflow changes its character from subsonic to supersonic).
For a given luminosity the choice of $T_{\ma c}$ does not affect the
results since a different temperature value essentially shifts the critical
point position leaving all physical quantities profiles almost unaffected.
We have shown in Paper I that
the characteristic wind number density, $n_{\ma c}$ 
(related to the number density at the critical point, as defined in Sect.~3)
 is a critical parameter, and the system is
very sensitive to its variations. Indeed, a solution for the nuclear wind equations is 
possible and characterized by a total electron scattering optical depth of the wind plasma 
$\simlt 1$ (see Paper~I) only over a rather narrow range of values for $n_{\ma c}$.
Variations of $n_{\ma c}$ within this range could in principle affect the BLR model and consequently 
might change the emission line properties; we discuss this issue in Sect.~5.

{\it  b) Parameters characterizing the nature of the stellar cluster orbiting around the
nucleus:} star number density distribution $\rho _*(r)$, 
mass-loss rate  and radius ($\dot M_*, R_*$) for each type of star (SG, RG, MS), 
and percentage of each type of stars present in the
cluster ($f_{\ma i}$ for stellar type ``i''). These parameters
enter the computation of stellar envelope properties, but also
contribute to determine the mass deposition function $f_{\rm m}(r)$
and hence influence the nuclear wind integration (see Sect.~3).
Some parameters have been  fixed
to their ``standard'' values.  This is the case for the star
percentages and radii which are set as
$f_{\ma{SG}}=10^{-4}$, $f_{\ma{RG}}=10^{-2}$, $f_{\ma{MS}}\sim 1$
and $(R_*)_{\ma{SG}}=100 R_{\odot}$, $(R_*)_{\ma{RG}}=10 R_{\odot}$,
$(R_*)_{\ma{MS}}=1 R_{\odot}$. For  others, namely $\rho _*(r)$ and
$(\dot M_*)_i$, different possibilities have been analysed. In particular
for stellar  mass loss rates the assumed values and their
 influence on the final results are described in the following sections.

\begin{figure}
\resizebox{\hsize}{!}{\includegraphics{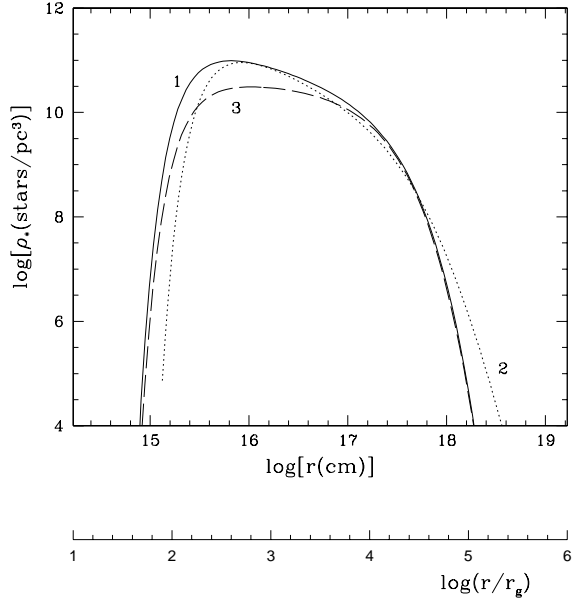}}
\caption{Star number density as a function of
the distance $r$ from the AGN nucleus. We also show the abscissa axis for the corresponding 
dimensionless distance $x=r/r_{\ma g}$  for the case of a
central black hole of mass $5.6\times 10^7~ M_{\odot}$, that we have chosen to associate to 
a central luminosity $L=  10^{44}$~erg/s (see Paper~I). Label meaning is explained in the text.}
\label{star}
\end{figure}

As far as the star number  density distribution is concerned,
there is no observational information about its detailed profile
in regions so close to the central nucleus, the only
possible indications come from theoretical work.
Therefore, we have chosen $\rho _*(r)$ profiles in accordance with
two main suggestions derived from the results of David et al.
 (\cite {david87a}, \cite {david87b}) analyses of a star cluster
evolution around a black hole. 
The first suggestion regards  
the peak value for the star number density in an evolved cluster,
that turns out to be  around $10^{11}$ stars/pc$^3$. The second
one is that  an evolved star distribution is concentrated in a range
of distances from $10^{14}$~cm up to $10^{18}$~cm
from the central black hole and steeply decreases both for shorter and for
longer distances.
We have analysed several different $\rho _*(r)$ profiles obeying these criteria,
and present here the results for 
 the stellar distribution profiles that  seem to best represent our working
scenario.  The selected curves are shown in Fig.\ref{star}.
In this figure the curve labeled ``1" is  representative of  the class of star
density distribution  which
gives the best results for our model. All the results presented in the following
sections refer to this particular choice of $\rho _*(r)$. 
The other two curves have been illustrated since  
an analysis of the consequent results 
allows us to highlight separately 
the effects of increasing $\rho _*(r)$
in the external region (curve ``2") and those of decreasing
$\rho _*(r)$ in the intermediate region (curve ``3").

Possible differences in the stellar distribution profiles
 in the region 
$r < 2\times 10^{15}$~cm
have proved to be 
unimportant in the present context, owing to the fact that
 our integration of the nuclear wind equations
stops at around that distance.
We elected to retain the same profiles $\rho_*(r)$ 
as shown in Fig.~\ref{star}, when analyzing configurations with different central black hole 
masses (that we associate with a different central luminosity, see Paper~I), thus implying  different 
values for $r_{\ma g}$, and, consequently  different 
resulting profiles as functions of the dimensionless distance $x\equiv r/r_{\ma g}$. 

{\it c) Parameters defining the physical properties
of the stellar envelopes:}
 stellar wind velocity profile (see Eq.(\ref{velocR})) and
 temperature value ($v_{0i}$,
$\alpha_i,  (T_*)_i$), for each type of star (``i"= SG, RG, MS).
This group of parameters only influences the properties of the
line-emitting regions.

 In summary, once we have chosen the luminosity of the AGN
for which we want to build up a model,
the free parameters we can work with turn out to be the following quantities:
 $n_c$, $\rho _*(r)$, $(\dot M_*)_i$, $v_{0i}$,
$\alpha_i$ and $  (T_*)_i$. The ``observational
requirements'' on the BLR described in the following subsection dictate further limitations
to the number of parameters that significantly influence the model.

\medskip

\subsection{BLR parameters and observational requirements }

Not every combination of the above listed free parameters
can reproduce the observed BLR general properties. Observative and
interpretative work has shown that  definite ranges of parameters
and/or  quite specific features are common to the BLRs of most AGNs. 
In this section these
observational requirements are ``translated'' in terms of our model
variables.

 From the observational and interpretative points of
view, our spherical stellar atmospheres can
contribute to the BLR emission only if $R_{\ma{ext}}(r)$ and $n_*(r)$ satisfy
the following conditions:

~a) $R_{\ma{ext}}(r)$ $\geq \Delta R$ $= 10^{23}U/n_*$ cm,
since  photoionization equilibrium requires that emission lines are generated
in  shells of thickness
$\Delta R$, which is at least $\Delta R= 10^{23}U/n_*$ cm
(see eg. Peterson \cite{peterson97}).
The definition of an emitting shell then allows an estimate of the mean density of
the emitting plasma in the shell itself. Since the density inside the stellar 
expanding envelope
varies as shown in Eq.(\ref{nenv}), the mean density can be evaluated as  
\beq
\hat n_*(r)= \left({\dot M_*\over 4\pi~m_{\rm H}v_0 R_*^2}\right)
\int^{R_{\ma ext}}_{R_{\rm ext}- \Delta R}
\left ( {R \over R_*} \right )^{-(\alpha+2)} {dR \over \Delta R},
\label{meand}
\eneq
where both the integration limits  and possibly 
the mass loss rate depend on $r$. In terms of the quantity just defined,
the other conditions a BLR emitting plasma must satisfy read:

~b) $\hat  n_*(r) \geq 10^{8}$ cm$^{-3}$,
 since broad forbidden lines
are not observed (Netzer \cite{netzer90}, Peterson \cite{peterson97});

~c) if the line width is due to thermal effects only,
it must be  $ \hat n_*(r) < 1\times 10^{12}$ cm$^{-3}$
to avoid line thermalization, {\it i.e.}, for line emission to be
effectively significant  (see Rees, Netzer \& Ferland \cite{rees89},
Kaspi \& Netzer \cite{kaspi}, Korista \& Goad \cite{korista00}).

In our model
  $R_{\rm ext}(r)$ and $\hat n_*(r)$ will attain different
values at different distances from the central black hole, hence
 conditions a), b) and c) will
be satisfied only in a specific range of distances.
We call $r_1$ and $r_2$ the distances which delimit this interval.
Stellar envelopes satisfying points a) to c) contribute to
build up  the total covering factor. 
Here  we define
\begin{equation}
C_{\ma i}(r)=\int^{r}_{(r_1)_{\ma i}} \pi
 (R_{\ma{ext}})^2_{\ma i} f_{\ma i}\rho_* dr,
 \label{cv}
\end{equation}
as the covering factor accumulated at distance $r$ from the central black hole 
due to contributions from  star type `i''
(``i''= SG, RG, MS), obtained using $R_{\ma {ext}}(r)$
computed for that specific star type.
To obtain the total contribution  to the covering factor due to  a given stellar type, 
in Eq.({\ref {cv}}) we have to integrate 
up to $r=r_2$, i.e. up to the external boundary of the interval defining the 
broad line contributing region for each stellar type, that is  we have to compute
$C_{\ma i}(r_2)$ for ``i''= MS,RG,SG.
Finally, we define the total covering factor $C_{\ma{tot}}$ for a given BLR model
as the sum of the total contributions, that is
$$
C_{\ma{tot}}\equiv C_{\ma{MS}}(r_2) + C_{\ma{RG}}(r_2) + C_{\ma{SG}}(r_2).
$$

Our model  can now
be compared with observations, testing  the following points:

1) reasonable values should be recovered for
the covering factor $C_{\ma{tot}}$, whose estimates range from about $0.05$
up to $0.25$ (Baldwin \cite {Baldwin97});

2) the typical distance  characterizing the BLR,
namely  $r_{\rm BLR} $ (see Sect.~1)
 must be contained in the intervals [$r_1, r_2$]$_{\rm i}$ relative
to the stellar types contributing to the definition of the value of $C_{\ma{tot}}$;

3) the covering factor relative to ``broad'' forbidden emission lines,
$C_{\rm forb}$, must be vanishing. In our model, we define
\begin{equation}
{ C_{\ma{forb}}= \sum _{\rm i} (C_{\ma{forb}}) _{\ma i}
=\sum_{\ma i} \int^{\infty}_{r_0} \pi (\hat R_{\ma{ext}})_{\ma i}^2
 f_{\ma i}\rho_* dr,}
 \label{cvforb} 
\end{equation}
where
$$(\hat R_{\ma{ext}}) _{\ma i}= \cases{
(R_{\ma{ext}})_{\ma i}~~~{\rm if}~~~ n_*[(R_{\ma {ext}})_{\ma i}] <
  10^8~{\rm cm}^{-3}\cr
     0 ~~~~~~~~~~{\rm if}~~~n_*[(R_{\ma{ext}})_{\ma i}] \geq 10^8~{\rm cm}^{-3}\cr}$$
(see Peterson \cite{peterson97}, Krolik \cite{krolik99}), therefore, a contribution to
$C_{\ma{forb}}$ comes only from  those envelopes 
whose photoionized shell interior to the confining 
bow shock has a density appropriate for forbidden line emission.

4) the ionization parameter should be in in the range 0.01-1.
Assuming $<h\nu> =2.7$~Ryd as in AN1, and with $L$ in erg/sec, $\hat n_*$ in cm$^{-3}$
and $r$ in cm,
 the expression for $U$ shown in the Sect.~1 becomes
\begin{equation}
U(r)=4.5\times 10^{-2} {L \over \hat n_*r^2}.
\label{uc}
\end{equation}

The above general requirements allow us to
 evaluate the validity of our model and to restrict the range in which
free parameters described in the above subsection can vary.

\medskip

{\section {Model results}}

In Paper~I, we have defined a model for an AGN nuclear wind, that turned out to be
essentially determined by the choice of the central radiation field luminosity, $L$, 
of $n_c$, the characteristic number density of the wind plasma 
(see Sect.~4.1),
and of the mass deposition function $f_{\ma m}(r)$ (see Sect.~3 and Paper~I).
As it is discussed in Sect.~3, the mass source function (and therefore ultimately
$f_{\ma m}(r)$) for the nuclear wind depends on the choice of both 
the star number density distribution of the central cluster, $\rho_*(r)$, and 
the star cluster composition and properties (such as mass loss rates, radii and percentages of the
three different star types that we assume as components of the stellar population of the cluster;
see Sect.~4.1 for details).
As a consequence, the choice of all the cluster parameters mentioned above 
strongly influences the general properties of the solution for the nuclear wind as
well. These, in turn, enter the determination of the physical quantities characterizing 
the stellar envelopes that we consider to be the line emitting sources. 

The above described complex interplay between star and interstellar
medium properties makes it difficult to  predict our model results.
While it is obvious  that the validity of our model could be
proved only by the detailed  comparison of predicted and observed line profiles,
it seems of primary importance to understand which are the relevant parameters
in our picture.
Hence, due to  the large number of parameters and the complexity of the
model, 
a general analysis aimed at performing  
a  first selection among all the present parameters
is preferable, in order to  get 
a deeper insight into the flexibility of the model itself. 
Therefore, as a first step in the comparison  between observed
BLR features and  our model results, 
we have analysed in detail the influence that each free parameter has 
in the fulfillment of the observational requirements presented in Sect.~4.2.
The results of this analysis are presented hereafter. We refer
to Sect.~4.1 for the specific definition of the values chosen 
and fixed as ``standard'' for stellar radii and percentages in the cluster.

{\it Effects of changing the star density distribution $\rho _*$}.\\
As we have mentioned in Sect.~4.1, point {\it b)}, our system is very sensitive to the choice
of $\rho _*$. 
Indeed, the results obtained by using the three stellar distributions shown in Fig.~\ref{star} turn out 
to be significantly different.
The  ``small'' differences
among $\rho_*$ profiles are sufficient to exclude the  star 
distribution labeled with ``2''.
In fact, any stellar distribution $\rho _*(r)$ that is more extended in the ``external'' region (i.e.,
for $r\simgt 10^{18}$~cm) 
than the one characterized by profile ``1'')
can reproduce a covering factor in accordance with point 1) of Sect.~4.2, 
but also shows  a resulting non-negligible forbidden line covering factor, so that it cannot
satisfy point 3) of Sect.~4.2. On the other hand, for profiles  similar to the one 
labeled with ``3'', i.e. characterized by  a smaller
amount of stars, the forbidden covering factor  is negligible, but  $C_{\ma{tot}}$
turns out to be too low with respect to the range of values inferred from observations
if suitably high values of stellar mass loss rates are not used.
This is apparent from Eq.(\ref{cv}), taking into account that $R_{\ma {ext}}^2$
is generally proportional to the ratio $\dot M/v_0$ (see Eq.(\ref{bow})).
Accounting for the considerations above, in the following we 
have chosen to discuss the results pertaining to the
the star number density profile labeled with ``1" in Fig.~\ref{star}, 
as representative  of a possible
cluster star distribution for broad line emitting AGNs.\\
{\it ~~Effects of changing the nuclear wind density parameter $n_{\ma c}$.}\\
 As shown in Paper I (Eq.s (\ref{cont}) and (\ref{minp})), the AGN wind integration system
 depends on the quantity
$f_{\ma m}'/f_{\ma m}$, which can be easily derived
 as
 \vskip 0.3cm
\beq
{f_{\ma m}' \over f_{\ma m}}= {r^2 F(r) \over A_0 f_{\ma m}(r_0) + \int^r_{r_0}
r^2 F(r) dr}
\label{refe}
\eneq
where  it is 
$f_{\ma m}'=r^2 F(r)/A_0$ from 
Eq.(\ref{fr}). 
Here the parameters of the star cluster define 
the mass source function per unit solid angle, that is the function $F(r)$, through
Eq.(\ref{starden2}) 
and $f_{\ma m}(r_0)$ is an integration constant.
Once chosen a specific model, i.e. for fixed $F(r)$, the relative weight
of the terms $A_0 f_{\ma m}(r_0)$ and 
$\int^r_{r_0} r^2 F(r) dr$ 
defines the amount of of mass flux characterizing the 
nuclear wind at its base with respect to that fed by stars.

Hence, changing $n_{\ma c}$, which enters the definition of $A_0$,  changes the function 
$f_{\ma m}' / f_{\ma m}$
mainly in the ``internal'' region ($r \sim r_0$) where the term $A_0 f_{\ma m}(r_0)$ is
non-negligible with respect the integral term. Increasing $n_{\ma c}$,
the intrinsic mass flux of the nuclear wind
becomes more important with respect to the mass flux fed by stars 
and the electron scattering optical depth may become too large (as discussed in Paper I). 
For $r>>r_0$,   in the region where the 
BLR is located, the integral 
$\int^r_{r_0} r^2 F(r) dr$ 
is the predominant term in the mass flux definition and a change in the
critical density value  does not have a significant effect on the final result.
A numerical analysis confirms the present analytical discussion, therefore 
we conclude that, once $n_{\ma c}$ is chosen in the restricted range allowed by the nuclear wind
integration (as explained in Sect.~4.1 a), its changes do not affect the BLR properties
as derived from our model.
 
{\it Effects of changing the velocity profile of the stellar wind, i.e. the value
of $\alpha$, in Eq.(\ref{velocR})}. \\
Several integrations of our model, carried
on with different $\alpha$ values, have shown that the  specific velocity profile
 does not have a strong influence
on the results. 
However,  positive $\alpha$ values are generally preferable,
since the corresponding models are characterized by lower values of the forbidden line covering
factor with respect to those with negative $\alpha$ values.
Perhaps, the most straightforward way to understand this is to go along
the following line of reasoning.
When the stellar envelope extension is delimited by the bow shock
(which is the case under most of the conditions of physical interest),
the density of the envelope at its boundary, $n_*$, is
determined by the local density of the AGN nuclear wind plasma, since
$n_*\propto n_{\ma w}$, 
as it can be realized by inserting Eq.(\ref{bow}) in Eq.(\ref{next}). 
This implies that, going farther from the AGN central black hole,
$n_*$ decreases, like $n_{\ma w}(r)$ does.
Accounting for this condition in Eq.(\ref{next}), it is easy to see that, for a given $\dot M$,
$R_{\ma{ext}}$ increases with $r$, and  it  increases faster
for negative values of $\alpha$.
As a consequence of this last point, for negative $\alpha$ values non-negligible
contributions to the covering factor are expected to come from regions that are far enough
(in the computations that we present in the following subsections, this is generally
for $r\simgt 2-3\times 10^4r_{\ma g}$) from the center of the AGN, so that the stellar envelope 
density is  sufficiently low to be suitable for forbidden line emission. 

{\it Effects of changing the star envelope temperature}.
Values of the temperature in the range $1-3 \times 10^4$~K,
appropriate for BLR emission,
 have  been tested,  obtaining no significant change in the results.
This fact supports our simplifying assumption of
a constant temperature profile for the overall
stellar expanding envelope.

Taking into account all the above points, we conclude that in our
 BLR model only  a specific class of
star number density profiles meets the observational requirements, 
 while changes in the values of the
parameters $n_c$, $\alpha$ and $T_*$ 
do not have a relevant influence on the quality of the reproduction of BLR properties.
For this reason, in the following 
we illustrate the general results of our exploration
of  physically reasonable ranges of the parameters whose changes turned out from our 
analysis to significantly affect the results as for BLR global properties, namely 
stellar mass loss rates and expansion velocity of the stellar atmospheres (see end of 
Sect.~4.1). To do this, we have chosen to discuss separately three different choices for 
the definition of the mass loss rates in the envelopes of the different stellar types composing the
cluster, and to examine the effects of changing the expansion velocity of the envelopes 
on a significant quantity such as the resulting covering factor of the proposed BLR model.
Before entering the details of the discussion, we just want to remind that 
all the results that we show and comment in the following, unless otherwise specified, 
refer to a specific choice for 
\begin{description}
\item{(i)}~~~ the central source luminosity,  $L=10^{44}$~erg/sec, 
\item{(ii)}~~~ the black hole mass, $M_{\ma{BH}} = 5.6\times 10^7 M_{\odot}$
	   (thus defining $r_{\ma g}$),
\item{(iii)}~~~the envelope temperature, $T_*=2\times10^4$~K (see Sect.~4.2),
\item{(iv)}~~~the expansion velocity power law exponent, $\alpha = 0.75$ (see Sect.~4.2),
\item{(v)}~~~the star number density, $\rho_*(r)$, labeled ``1'' in Fig.~\ref{star}, 
\end{description}
where choices (iii), (iv) and (v) are justified by the above discussion. 

As for the stellar mass loss rates, we have chosen to discuss 

a) the simple case of ``standard'' mass loss rates for stellar winds of stars of a given 
type, as known from galactic studies (see below for the details and values,
Sect.~5.1);

b) a case in which we try to account for the possibility of ``enhanced'' mass loss 
of MS, RG, and SG stars, due to the very specific context in which they are set, namely a strong
X-ray illumination from the  central source in the AGN (see Sect.~5.2);  this would lead 
to an expected enhancement of mass loss rates increasing with decreasing $r$, 
with a behaviour following the radiation 
flux increase going inwards, that is $\propto r^{-2}$;

c) a third case in which we suppose the mass loss rates do again depend on the distance from 
the central black hole $r$, but with a flatter 
increase going inwards (Sect.~5.3). The physical mechanism responsible for this configuration
is not specified in this case, but we refer to Sect.~6 for a brief discussion of possible
processes.

 \medskip
 \medskip

{\subsection{Standard mass loss rates}}

The first reasonable step is that of assuming ``standard'' mass loss rates
for the stars in the cluster (see, for example, Lamers \& Cassinelli \cite{lamers}),
that is    
$$(\dot M_*)_{\ma{SG}} \simeq 10^{-6} M_{\odot}/{\rm yr}, $$
 $$(\dot M_*)_{\ma{RG}} \simeq 10^{-8} M_{\odot}/{\rm yr}, $$
$$(\dot M_*)_{\ma{MS}} \simeq 10^{-14} M_{\odot}/{\rm yr}.$$

With this choice for the mass loss rates of the stars in the central cluster, 
from Eq.~(\ref{fr}) it turns out that, in the inner region 
of the cluster (i.e., for $r\simlt 10^4r_{\ma g}$), the mass flux input in the nuclear wind is 
strongly dominated by collisional 
effects (i.e., mass loss from collisions between stars). On the contrary, farther from the AGN 
center there 
is a vast region in which predominance 
of the stellar wind mass loss sets on.
We then derive the corresponding AGN nuclear wind model, and, consistently, the main 
physical parameters of the BLR emitting envelopes.

For each stellar type these are: i)~the evaluation of the characteristic extension 
for the expanding envelope, 
$R_{\ma{ext}}$ (see Eq.~(\ref{ext})), and the physical mechanism defining it; ii)~the average 
density of the envelope
over the emitting region of the envelope itself, $\hat n_*$, (see Eq.~(\ref{meand})); iii)~ the resulting
ionization parameter $U$ (Eq.(\ref{uc})); iv)~the covering factor, as a function of the distance $r$ 
from the black hole in the interval $[r_1, r_2]$ defining the BLR (see Eq.~(\ref{cv})).
A typical behaviour (which, we want to stress, is generally common to the three cases for stellar mass loss 
rates that we present in this work) corresponds to an increasing envelope extension, 
and to a decreasing 
density with $r$ increasing. Of course, the covering factor $C_{\ma i}(r)$ is an increasing 
function of $r$ over the model BLR range of distances, since the contributions from the stars 
in each shell at a given distance from the black hole accumulate in the integral defined by
Eq.~(\ref{cv}). 

First, we want to focus our attention on the 
results we obtain from the model in terms of global parameters of the 
BLR, such as the integrated covering factor, defined as 
$C_{\ma i}(r_2)$ for each stellar type ``i''.
We briefly describe the $r$ dependence of the first three physical quantities 
mentioned above later on in the present Section and 
in Sect.~6.

\begin{figure}
\resizebox{\hsize}{!}{\includegraphics{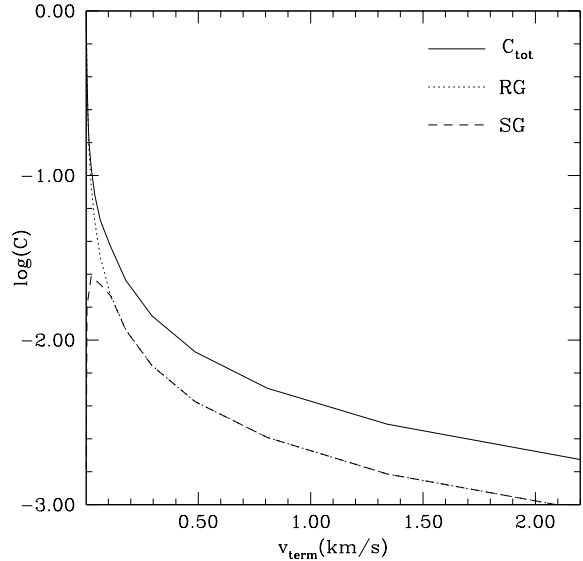}}
\caption{
The total value of the covering factor, $C_{\ma {tot}}$, defined as the sum of the 
contributions due to the envelopes of different stellar 
types, as well as the distinct contributions to this value from the various stellar types, 
are shown as  functions
of the terminal velocity  of expansion $v_{\ma{term}}$ (in km/s), characterizing the contributing 
envelopes. The present figure refers to the case for ``standard'' mass loss rates.
Notice that the contribution due to main sequence stars in this
``standard'' case does not even appear in the plot, since it is absolutely negligible
with respect to those due to RG and SG stars.}
\label{cvstan}
\end{figure}

The general outcome for this case is exemplified in Fig.~\ref{cvstan}, showing the resulting
covering factor  contributions (computed from Eq.~(\ref{cv}) with $r=r_2$) for the modeled BLR, as  functions of 
the terminal velocity $v_{\ma{term}}\equiv v_*(R_{\ma {ext}})$, characterizing the stellar envelopes 
that are assumed to contribute to the broad line emission. 
We also plot the sum of the contributions due to the different 
stellar types composing our model central cluster as $C_{\ma{tot}}$, again
as a function of $v_{\ma{term}}$. 
We obtain the variation of $v_{\ma{term}}$ for each stellar type,
by varying $(v_0)_{\ma i}$, which, in turn, also induces
variations in the extension of the envelope, at least when the confinement mechanism
for the envelope itself is the bow shock (see Sect.~2, Eqs.~(\ref{bow})
 and (\ref{rcomp})); changing this parameter is therefore significant to the 
determination of the  covering factor.
To give an idea of the range of $(v_0)_{\ma i}$, we report here  the values
corresponding to  the terminal velocities $v_{\ma{term}}=0.1$ and 2.1~km/s, which 
are close  to the extremes of the terminal velocity range plotted in Fig.~2 and
representative of ``low'' and ``high'' terminal velocities respectively.
In the present case, for the star types that do give contribution to the covering factor, 
the values of $v_0$ are
$(v_0)_{\ma {RG}}(v_{\ma{term}}=0.1~{\rm km/s})= 1.6\times 10^{-3}$~km/s, 
$(v_0)_{\ma {SG}}(v_{\ma{term}}=0.1~{\rm km/s})= 3.7\times 10^{-3}$~km/s, and 
$(v_0)_{\ma {RG}}(v_{\ma{term}}=2.1~{\rm km/s})= 1.8\times 10^{-2}$~km/s, 
$(v_0)_{\ma {SG}}(v_{\ma{term}}=2.1~{\rm km/s})= 4.2\times 10^{-2}$~km/s. 

It is important to note that
the plotted value of $C_{\ma{tot}}$ is the one that we obtain when we suppose that the 
three different star type envelopes are endowed with the same value of the terminal velocity, which 
is not necessarily the case. In fact, one can obtain different values for the total covering factor 
for the BLR model, with respect to those that we present in the plot, by assuming  different 
$v_{\ma{term}}$ values for each single species of stars, that is picking contributions 
from the various star types at different $v_{\ma{term}}$ along the curves defining the specific 
covering factor for a given stellar type. 
However, we want to stress that, 
due to the fact that the  profiles $C_{\ma i}(v_{\ma{term}})$ for the three stellar
types show a similar behaviour, which is generally decreasing 
for increasing $v_{\ma{term}}$ (apart from the very low velocity end of the range shown in the figures),
it turns out that for any chosen value of $v_{\ma{term}}$
the total covering factor we plot in Fig.~\ref{cvstan} 
is the maximum value that can be attained if we suppose that the different contributing stellar type 
envelopes are characterized by 
terminal velocity values that are possibly different, but in any case larger than the  one chosen. 
On the contrary, larger $C_{\ma{tot}}$ values can be obtained by allowing for different terminal
velocities of different stellar type envelopes, provided the $v_{\ma{term}}$ value of at least 
one of the contributing types is lower than the one chosen for the exemplification.
These are general considerations and they refer both to the present and to the other
two cases we discuss in the following subsections. 
 
\begin{figure}
\resizebox{\hsize}{!}{\includegraphics{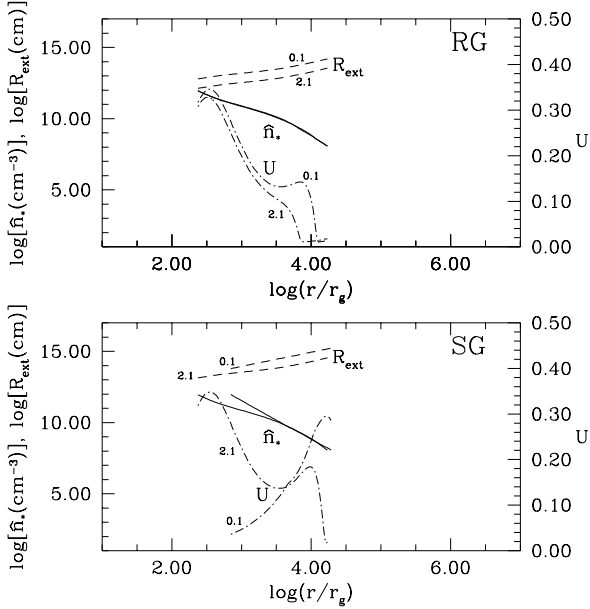}}
\caption{Physical parameters characterizing stellar envelopes contributing 
to the BLR as functions of the normalized distance $r/r_{\ma g}$ from the central black hole;
these are the mean density of the envelope, $\hat n_*(r)$, the estimate of the envelope
extension, $R_{\ma{ext}}(r)$, and the ionization parameter, $U(r)$.
The interval over which the physical quantities are plotted represents the ``BLR'' interval
$[r_1,r_2]$ as defined in Sect.~4.2; the labels 0.1 and 2.1 
identify the specific value (in km/s) of the terminal velocity parameter for which they are 
computed. The upper panel refers to RGs' envelopes, while the lower one shows the results
pertaining to the envelopes of SG stars, in the case of ``standard'' mass loss rates.}
\label{standfig}
\end{figure}

An inspection of the solutions for the BLR physical parameters shows that the bow shock mechanism
is in general the most efficient in confining the stellar envelopes and, consequently, determines their
extension for all the three different stellar types.  As a consequence, for this specific choice
of the stellar mass loss rates, the envelope radii for RG stars and SG stars turn out to scale so 
as to give essentially the same contribution to the total covering factor for the BLR.
As it is apparent from Fig.~\ref{cvstan}, this is no longer true for low velocities ($\simlt 0.12$~km/s),
when the contribution to the covering factor due to SG stars ``decouples'' from that
of RGs. The physical reason for this is in the fact that at these low velocities 
the most efficient confining mechanism for SGs all through the BLR extension becomes tidal 
disruption of the envelope beyond a given radius (see Eq.~(\ref{rtid})), whereas for RGs the bow shock
mechanism is still dominant.
 
To illustrate the behaviour of the BLR physical quantities and the characteristic extension of our 
BLR model (i.e., the $[r_1,r_2]$ interval
over which conditions a) to c) of Sect.~4.2 for the emitting gas structures are fulfilled), we 
have chosen to plot in the two panels of Fig.~\ref{standfig} (each one referring to one of the two stellar
types that give substantial contribution to the broad line formation, namely RGs and SGs in this case) 
the three quantities $R_{\ma{ext}}$, $\hat n_*$ and $U$. These are functions
of the distance from the central black hole $r$, and we plot two curves for each quantity, 
corresponding to the two 
representative values of the expansion velocity 
parameter $v_{\ma{term}}$ previously defined, namely 
$v_{\ma{term}}=0.1$~km/s and $v_{\ma{term}}= 2.1$~km/s, 
close to the extremes of the velocity range plotted in Fig.~\ref{cvstan}.
For the sake of clarity, we note that the $v_{\ma{term}}$ values chosen
refer to those attained for envelopes at $r=r_2$, that is at the external border of the broad line emitting region.
For each BLR physical parameter the curves obtained for intermediate terminal velocity 
values typically lie in the region  between the 0.1 and 2.1 curves.  Therefore, each of the 
two panels of Fig.~\ref{standfig} 
essentially shows the ranges of values for the BLR parameters for the models corresponding 
to the present choice of mass loss rates for the stars of the central cluster.
It is interesting to notice that for RGs the extension of the BLR  contributing region is independent of the 
terminal velocity value, whereas this is not the case for SGs. 
This is again due to the fact that for RGs the envelope radius is defined by the bow-shock mechanism, 
whereas at low $v_{\ma{term}}$ it becomes $(R_{\ma{ext}})_{\ma{SG}}=(R_{\ma{tidal}})_{\ma{SG}}$.
Indeed, it is easy to see that, when the bow-shock is the confining physical mechanism for 
the stellar envelope, and the expansion velocity is subsonic, $n_*(r)$ (see Eqs.~(6) and (7)) 
is essentially independent 
of the stellar wind velocity itself. Moreover, when the envelope gas temperature is chosen to be
 the same for the three stellar types, $n_*(r)$ is independent of the stellar type as well.
The same considerations do hold for the average density in the emitting portion of the envelope,
$\hat n_*$ (see Eq.~(\ref{meand})), which is the quantity that we plot in Fig.~\ref{standfig}, since in general 
it turns out that  $\hat n_*=n_*$. In fact, our computations have shown that the emitting shell
of an envelope is characterized by a size $\Delta R<<R_{\ma {ext}}$, so that, performing an
analytical integration of Eq.~(\ref{meand}) and using a first order approximation in 
$\Delta R/R_{\ma{ext}}(<<1)$, one ends up with $\hat n_*=n_*$.
As a result, the limiting conditions on $\hat n_*$ for line emission (points b) and c)
of Sect.~4.2)
are met at the same distance from the black hole, $r$, independently of the chosen value for the 
velocity parameter, since (see Eqs.~(\ref{ptoeq}) and (\ref{next})) they only depend on  the nuclear wind physical 
properties,
namely $n_{\ma w}(r)$, the nuclear wind density, $T_{\ma w}(r)$, its temperature profile, and 
on the velocity defining the ram pressure at the envelope boundary (see Sect.~2).
We want to stress here that these same considerations do hold 
independently of the choice of mass loss rates, therefore they similarly apply to the cases 
discussed in the next subsections as well.
For completeness, in Fig.~4 we show the behaviour of the AGN wind physical quantities 
$n_{\ma w}(r)$ and
$T_{\ma w}(r)$, as obtained from the nuclear wind system resolution for this specific choice
of the mass deposition. For this density profile the resulting total optical depth to
electron scattering for the nuclear wind plasma is $\tau_{\ma T}\simeq 0.12$.
Note that, for comparison, we have also plotted in Fig.~4 the $n_{\ma w}$ density profile (dotted curve) resulting 
from the computation of the solution for the nuclear wind with a different mass deposition, that is the one that 
will be extensively discussed  in Sect.~5.3, and corresponds to our ``most extreme'' (i.e., strongest 
$\dot M_{\ma {inp}}$) choice for 
mass deposition in the present work; the temperature profile corresponding to this latter case is not shown, since 
the differences with the dashed curve of Fig.~4 are qualitatively irrelevant.
\begin{figure}
\resizebox{\hsize}{!}{\includegraphics{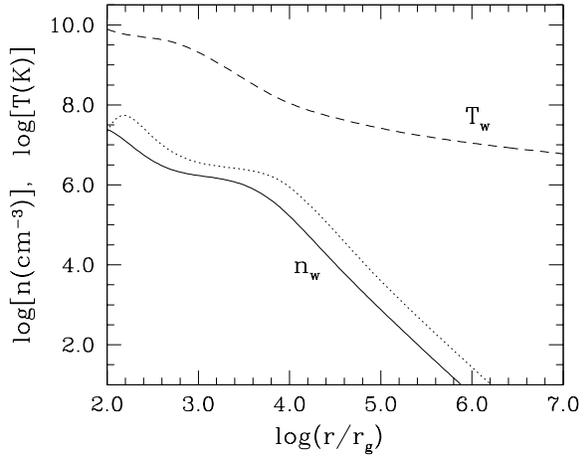}}
\caption{Temperature (dashed line) and number density (continuous line) profiles
for the AGN nuclear wind solution obtained  for the case of mass deposition defined by 
the choice of ``standard'' mass loss rates for the cluster stars. For comparison, we also plot as the dotted curve
the ``most extreme'' density profile that we have obtained, namely the one resulting from the solution for the 
nuclear wind with the mass deposition defined by the choice of mass loss rates described as case c) in Sect.~5 
and discussed in detail in forthcoming  Sect.~5.3.} 
\label{AGNstand}
\end{figure}

Going back to the limiting conditions (points b) and c) of Sect.~4.2) for the BLR
contributing envelope density $\hat n_*$, 
for the present case they define the interval $[r_1,r_2]_{\ma{RG}}\simeq [237 r_{\ma g},1.7\times 10^4 r_{\ma g}]
= [3.95\times 10^{15}$~cm,$2.8\times 10^{17}$~cm], thus including any reasonable estimate 
of the characteristic size $r_{\ma{BLR}}$ for our chosen value of $L$.

From Fig.~\ref{cvstan}, it is apparent that, within the ``standard'' mass loss rate case defined above,
significant covering factors can be recovered only when the terminal velocities of expansion
for the stellar envelopes are quite low, as compared with the typical sonic speed for the envelope gas
(for $T_*\simeq 2.\times 10^4$~K, we get $c_{\ma s}\simeq 21$~km/s). To obtain
$C_{\ma{tot}}\simeq 0.1$, $v_{\ma{term}}$ should be around 0.027~km/s.
Indeed, this is a very low value for the expansion velocity of a ``normal'' stellar
expanding atmosphere. However, we would like to mention that similarly low values have been 
presented in AN2 (\cite{alexander97}) 
in their own attempt to model the BLR structure as an ensemble of expanding stellar envelopes.
In fact, transposing their  notation to ours, these authors (AN2) assume $\alpha = -0.5$, 
a decreasing function of distance from the star surface for the stellar wind velocity,
and they choose to analyze models with $v_0\simeq 0.1$~km/s, where the value of 
$v_0$ is, in their case, the peak value of the stellar wind velocity.

We want to note that the present case of ``standard'' mass loss rates seems to be not 
very much relevant to the construction of a 
physical BLR model for another reason, namely the fact that the calculation of the resulting
covering factor for forbidden lines, which are actually not seen in the BLR, gives uncomfortable 
results. In fact, the global value of $C_{\ma{forb}}$ (as defined in Eq.~(\ref{cvforb})) turns out to reach
a significant fraction of the obtained $C_{\ma{tot}}$. For example, when $C_{\ma{tot}}\simeq 0.1$, 
we get for the present case $C_{\ma {forb}}\simeq 0.037$, which is not negligible.
This is related to the characteristics of the nuclear wind solution for the present case, in which
not much mass is input in the nuclear wind as a consequence of stellar mass loss, and, therefore,
the nuclear wind plasma density $n_{\ma w}$ is rather low in the ``external'' (i.e., at $r\simgt
10~r_{\ma{BLR}}$) region; this allows for  the formation of larger expanding envelopes, characterized
by density values that are sufficiently low to contribute to the increase of the covering factor
for forbidden lines (see the trends shown in Fig.~\ref{standfig} for the envelope 
physical parameters). 

As a final remark, we note that the total number of stars contributing to the BLR, that is essentially
the number of  RGs in the interval of distances $[r_1,r_2]$ for this case, turns out to be 
$~2\times10^5$. The total number of stars present within the bounds of the BLR for this
case is therefore $\sim 2\times 10^7$.
\medskip

{\subsection{Enhanced mass loss rates}}

It is by now well known that in a close binary system, intense X-ray illumination  from the 
primary star can have noticeable effects on the secondary star.
The best studied example is that of HZ Her (in the system HZ Her/Her X-1),
whose illuminated face originates an ``X-ray-excited'' wind
(see e.g. Hameury et al. \cite{hameury93}).
The effect is related to the intensity of the X-ray
flux at the stellar surface, and from this intensity it is possible to estimate, 
through an efficiency factor, $\eta$,
the fraction of luminosity  that gets transformed in kinetic power of the wind.
By physical analogy, we suppose that the stars of the AGN central cluster
may suffer from a similar effect, since they are strongly illuminated by the  hard 
radiation of the central radiation field of the AGN itself (see, for example, 
Taylor \cite{taylor}).
We can therefore try  an analogous (see  Hameury et al. \cite{hameury93}) estimate of the 
transfer of kinetic power into the stellar wind of stars belonging to our 
central cluster in a radio-quiet AGN, from the X-ray luminosity, $L_{\ma X}$, of the radiation 
field of the AGN central source, as follows:
\beq
\eta ~L_{\ma X} \left ({R_* \over 2 r}\right )^2 \simeq v_*^2 \dot M_*.
\label{enhanc}
\eneq
The value of $\eta$ computed for Her X-1 is $3\times 10^{-3}$
and it must be in any case $\eta <0.1$ (Hameury et al. \cite{hameury93}).
Taking into account that $L_{\ma X}< L$ , we assume $\eta L_{\ma X} \simeq 10^{-3}L$.
We notice here 
that, in order to produce BLR models that meet the general observational requirements 
discussed in Sect.~4.2, the resulting velocities for the expanding stellar envelopes 
are significantly lower than the sonic speed (by an order of magnitude or even more).
Therefore, it seems appropriate to derive an estimate of the ``enhanced'' mass loss
rate for any given type of stars in the cluster with a $v_*$ value close to the ones 
that correspond to models that better fulfill the observational conditions on the BLR.
In fact, we choose to use $v_*\simeq 3$~km/s 
to infer an order of magnitude value  for the induced mass loss
rate of a star in the central cluster of the AGN from relation~(\ref{enhanc}).
For $L=10^{44}$ erg/sec,
we thus find an estimate for the  enhanced mass loss rates

\beq
(\dot M_*)_{\ma {MS}} \simeq 8\times 10^{-2} (r/ r_{\ma g})^{-2} M_{\odot}/{\rm yr},\\
\label{sms}
\eneq
\beqar
(\dot M_*)_{\ma {RG}} \simeq 8 (r/ r_{\rm g})^{-2} M_{\odot}/{\rm yr},\\
\label{rg}
& & \nonumber\\
(\dot M_*)_{\ma {SG}} \simeq 8\times 10^2 (r/ r_{\rm g})^{-2} M_{\odot}/{\rm yr}.
\label{sg}
\eneqar

We have computed models 
using  for the three types of stars considered the 
mass-loss rates given by Eqs.~(\ref{sms}) to (\ref{sg}) for $r$ such that those relations give values 
larger than the ``standard'' ones defined in Sect.~5.1, and maintaining, on the contrary,
those same ``standard'' values when the equations above would determine smaller values. 
At $r\sim r_{\ma {BLR}}$ the above estimates would give $(\dot M_*)_{\ma {MS}} \simeq 2.4\times 10^{-8} 
 M_{\odot}/{\rm yr}$, 
$(\dot M_*)_{\ma {RG}} \simeq 2.4\times 10^{-6} 
 M_{\odot}/{\rm yr}$, and
$(\dot M_*)_{\ma {SG}} \simeq 2.4\times 10^{-4} 
 M_{\odot}/{\rm yr}$.
These evaluations point out an interesting consequence of this enhancement mechanism for 
stellar mass loss rates. In fact, at distances comparable to the  characteristic 
BLR radius, this process 
does change substantially the ``intrinsic''
red giant and supergiant mass loss rates, by a factor $\simgt 10^2$,  
but, in the specific environment considered 
here, it has much more influence on the effective mass loss rate
of main sequence stars, that turns out to be increased by more than a factor $10^6$ at these 
distances, relevant to the BLR.
\begin{figure}
\resizebox{\hsize}{!}{\includegraphics{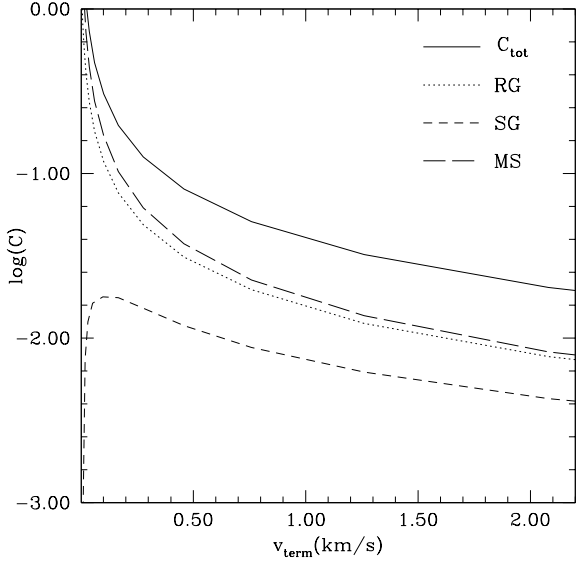}}
\caption{
The total value of the covering factor, $C_{\ma {tot}}$, defined as the sum of the 
contributions due to the envelopes of different stellar 
types, as well as the distinct contributions to this value from the various stellar types, 
are shown as  functions
of the terminal velocity  of expansion $v_{\ma{term}}$ (in km/s), characterizing the contributing envelopes. 
The present figure refers to the case for ``enhanced'' mass loss rates, and,
differently from the case shown in Fig.~\ref{cvstan}, the contribution due to main sequence stars is quite 
significant, and actually predominant with respect to RG and SG contributions.}
\label{cvsquare}
\end{figure}
\begin{figure}
\resizebox{\hsize}{!}{\includegraphics{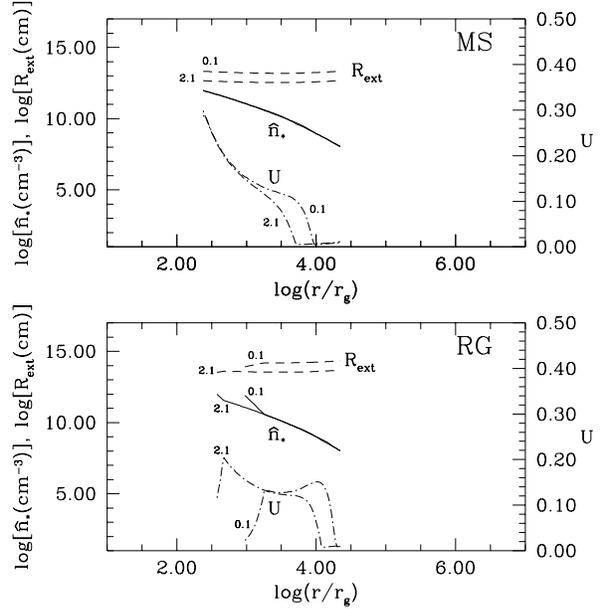}}
\caption{Physical parameters characterizing stellar envelopes contributing 
to the BLR as functions of the normalized distance $r/r_{\ma g}$ from the central black hole;
these are the mean density of the envelope, $\hat n_*(r)$, the estimate of the envelope
extension, $R_{\ma{ext}}(r)$, and the ionization parameter, $U(r)$.
The interval over which the physical quantities are plotted represents the ``BLR'' interval
$[r_1,r_2]$ as defined in Sect.~4.2; the labels 0.1 and 2.1 
identify the specific value (in km/s) of the terminal velocity parameter for which they are 
computed. 
The upper panel refers to main sequence star envelopes, while the lower one shows the results
pertaining to the envelopes of RG stars, in the case of ``enhanced'' mass loss rates.}
\label{squarefig}
\end{figure}
Therefore, the first important difference with respect to the ``standard'' mass loss rate 
case presented in Sect.~5.1 is  that the contribution of MS stars to the total covering factor
is no longer negligible for any value of the expansion velocity of the emitting envelopes. 
This is immediately apparent at a glance from Fig.~\ref{cvsquare}, which is the analogous 
to Fig.~\ref{cvstan} discussed
in the previous Sect.~5.1 for the case of ``standard'' mass loss rates.
The plot in Fig.~\ref{cvsquare} shows that the strong increase in mass loss rate for MS stars produces 
as a consequence a significant contribution to the covering factor. In fact, the extension 
of the emitting stellar envelopes of MS stars is determined by the bow shock mechanism
for basically any expansion velocity and the envelope radius is determined
by Eq.~(\ref{bow}); therefore, main sequence star envelopes turn out to be much larger than in the 
case of Sect.~5.1, due to the strong increase in $(\dot M_*)_{\ma{MS}}$.
Again, the number density $n_{\ma w}(r)$ and temperature $T_{\ma w}(r)$
of the nuclear wind are computed as the consistent solution of the nuclear wind system of equations 
with the present mass deposition choice.
In this case, the total optical depth to
electron scattering for the nuclear wind plasma turns out to be $\tau_{\ma T}\simeq 0.15$.

It is interesting to note that, due to the $r^{-2}$ decrease in $(\dot M_*)_{\ma i}$
moving outwards from the central radiation source, the MS star envelope 
extension, $R_{\ma{ext}}$, shows an extremely slow increase (if not a similarly extremely slow decrease 
locally) with increasing $r$. This can be seen directly from Fig.~\ref{squarefig}, similar to 
Fig.~\ref{standfig} for the ``standard'' mass loss case discussed in Sect.~5.1 (and we refer
to that Section for a detailed explanation of the notation and of the choice of 
the curves plotted); only, in the present case,
we show the results referring to MS stars and RG stars in the upper and lower 
panel respectively, since these two types of stars are the ones that mostly contribute to the 
present BLR model. 
As for RG envelopes, their extension is mostly bow shock determined, apart from the very inner 
shells of the BLR, in which the envelope radius is ``tidally'' defined; 
this ``tidally dominated'' 
portion of the RGs contribution to the BLR slowly extends to larger $r$ distances with 
decreasing the expansion velocity parameter. We note that this does not influence the behaviour
of the curve $C_{\ma{RG}}(v_{\ma{term}})$, since the portion of BLR over which the envelope
radius turns out to be ``tidal'' is in any case not much extended in $r$ and, especially, 
it is restricted to the inner region of the BLR contributing interval, whereas most of the 
contribution to the integrated covering factor $C_{\ma{RG}}(r_2)$ accumulates at larger values 
of the distance $r$, in the ``bow shock dominated'' portion of the interval
$[r_1,r_2]_{\ma{RG}}$. Analogously to the MS envelopes' case, in the 
``bow shock dominated'' region the envelope radius of RG expanding atmospheres is more or less 
constant with $r$, and this can be seen from the lower panel of Fig.~\ref{squarefig}.
The resulting interval of distances defining the extension of the MS stars contribution to the BLR is    
$[r_1,r_2]_{\ma{MS}}\simeq[239r_{\ma g},2.2\times 10^4r_{\ma g}]=[4.\times 10^{15}$cm,$3.7\times 10^{17}$cm].
RG stars contribute in a region $[r_1,r_2]_{\ma {RG}}$ whose extension shows a dependence
on the velocity parameter $v_{\ma{term}}$, as it can be seen from Fig.~\ref{squarefig}, but it does anyway 
contain the estimated value of $r_{\ma{BLR}}$, independently of $v_{\ma{term}}$ itself.
As for the mean envelope density $\hat n_*$, for the conditions in which the envelope radius is 
$R_{\ma{ext}}= R_{\ma{bow}}$, we refer to the considerations presented in
Sect.~5.1, that still hold. 

From Fig.~\ref{cvsquare} it is apparent that, at any value of the expansion velocity, the contribution 
to the total covering factor due to MS stars is dominant, but comparable in order of magnitude
with the RG contribution. On the contrary, the contribution of SG envelopes  turns out to be 
much less significant in this case. An analysis of the SG envelope extension as a function of 
$r$ shows that 
when the velocity parameter $v_{\ma{term}}$ becomes very small ($< 0.2$~km/s), 
the dominant confinement mechanism for the envelopes becomes generally different from the 
bow-shock one, similarly to what was discussed for the ``standard'' mass loss case 
in Sect.~5.1, thus leading to the change in slope of the curve $C_{\ma {SG}}=C_{\ma
{SG}}(v_{\ma{term}})$ observed in Fig.~\ref{cvsquare}. 
As for the definition of the range of values for the velocity parameter $(v_0)_{\ma i}$
for the present case, we report representative values that are analogous to those 
defined in Sect.~5.1 for the ``standard'' mass loss rate case, namely
$(v_0)_{\ma {MS}}(v_{\ma{term}}=0.1~{\rm km/s})= 1.4\times 10^{-3}$~km/s, 
$(v_0)_{\ma {RG}}(v_{\ma{term}}=0.1~{\rm km/s})= 1.4\times 10^{-3}$~km/s, 
$(v_0)_{\ma {SG}}(v_{\ma{term}}=0.1~{\rm km/s})= 1.4\times 10^{-3}$~km/s, and 
$(v_0)_{\ma {MS}}(v_{\ma{term}}=2.1~{\rm km/s})= 8.8\times 10^{-2}$~km/s, 
$(v_0)_{\ma {RG}}(v_{\ma{term}}=2.1~{\rm km/s})= 8.8\times 10^{-2}$~km/s, 
$(v_0)_{\ma {SG}}(v_{\ma{term}}=2.1~{\rm km/s})= 8.8\times 10^{-2}$~km/s.
Notice that here we have given the values for main sequence stars as well, since 
they do contribute to the covering factor 
in this specific case. Also, the values we report are exactly the same independently of the 
stellar type and this is due to the specific choice of the mass loss rates for the 
three stellar types, which turn out to give $(\dot M_*/R^2_*)_{\ma {MS}}=(\dot M_*/R^2_*)_{\ma {RG}}
=(\dot M_*/R^2_*)_{\ma {SG}}$, thus resulting in the same $(R_{\ma{ext}}/R_*)$ value at the BLR 
external border, defined by the condition $n_*=10^8$~cm$^{-3}$ for any stellar type (see Eq.~(\ref{nenv})).

It is easy to see that in the present case, we can build models corresponding
to values around 0.1 for the total covering factor for envelope expansion velocities significantly
larger than what results in the case of ``standard'' mass loss rates. In fact, 
we obtain 
$C_{\ma{tot}}\simeq 0.1$ for $v_{\ma{term}}\sim 0.4$~km/s, which is more than one order of magnitude
larger than the result obtained in the case discussed in Sect.~5.1. However, this 
velocities still are in a strongly subsonic range of values.
Another improvement, with respect to the global results of the first case (discussed in Sect.~5.1),
is the fact that for the present choice of stellar mass loss rates, the ratio 
$C_{\ma{forb}}/C_{\ma{tot}}$ turns out to be quite smaller, typically ranging 
between 0.02 and 0.04 for the range of velocity parameter shown in Fig.~\ref{cvsquare}, thus setting
the  value of $C_{\ma{forb}}$ significantly close to ``vanishing''.

Finally, we want to stress another important difference with respect to the case of the 
standard mass loss model,
namely the fact that in the present scenario we 
have a much larger number of distinct emitting units constituting the BLR structure, since we 
can include all  main sequence
stars of the central cluster characterized by distances $r$ from the central black hole that fall 
in the BLR  contributing interval $[r_1,r_2]_{\ma{MS}}$. For example, for models giving 
$C_{\ma{tot}} \simeq 0.1$, we get an estimate of the total emitting units in the 
BLR $(N_*)_{\ma{tot}} \simeq 2.7\times10^7$. 

\medskip

{\subsection{``Intermediate'' mass loss rates}}
 
As a third case for stellar mass loss rates, we have chosen 
another set of $r$-dependent $(\dot M_*)_{\ma i}$. In this case, however, we do not 
specify the 
physical motivation lying beneath the choice, like in Sect.~5.2, and we
only rely on the fact that other mechanisms could be responsible of generating a
condition in which mass loss rates of stars in the central cluster of an AGN
could be altered (and enhanced) with respect to the ``standard'' 
values (see Sect.~6 for a brief discussion). 
This possibility has been taken
into account also in previous analyses of BLR models based on stellar envelopes as individual
contributors to the line emission.
We refer, for instance, to the work of Scoville and Norman (\cite{scoville}), in which the
authors consider giant star envelopes in a central cluster as  emitting units; 
they suppose that for the giant stars an ``induced'' mass loss, that turns out to depend on 
the distance $r$, is generated by some unspecified external heating from the central source.
Therefore, our choice in the present Section is similar to the one of Scoville and 
Norman (\cite{scoville}),  since 
we do not analyse the  physical mechanism responsible for affecting the stellar mass
loss rates.

We have thus parameterized the mass loss rates that we take into account in the present 
Section as follows:
\beq
(\dot M_*)_{\ma {MS}} = 9\times 10^{-10} \left({r\over 10^6 r_{\ma g}}\right)^{-\beta} 
M_{\odot}/{\rm yr},\\
\label{sims}
\eneq
\beqar
(\dot M_*)_{\ma {RG}} = 3\times 10^{-7} \left({r\over 10^6 r_{\ma g}}\right)^{-\beta} 
M_{\odot}/{\rm yr},\\
\label{irg}
& & \nonumber\\
(\dot M_*)_{\ma {SG}} = 3\times 10^{-6} \left({r\over 10^6 r_{\ma g}}\right)^{-\beta} 
M_{\odot}/{\rm yr},
\label{isg}
\eneqar
where we have chosen 
$\beta = 0.3$. 
For any type of star, the equations above define the mass loss rate that we use for the models we 
discuss in this Section 
for $r$ up to the distance at which the value of the mass loss rate itself (which is a decreasing 
function of $r$) reaches the ``standard'' value mentioned at the beginning of
Sect.~5.1, whereas for larger distances we discard Eqs.~(\ref{sims})-(\ref{isg})
 and we assume that the mass 
loss rate is constant and given by the ``standard'' values of Sect.~5.1.
As for the nuclear wind number density and temperature profiles ($n_{\ma w}(r)$ and $T_{\ma w}(r)$)
derived from the nuclear wind solution referring to the present case for mass deposition, 
we remind that $n_{\ma w}(r)$ is shown in Fig.~4 as the dotted curve. The nuclear wind temperature profile
shows no significant differences with respect to the one shown in Fig.~4, from a qualitative point of view. The 
resulting total optical depth to electron scattering for the nuclear wind plasma is here $\tau_{\ma T}\simeq
0.4$.

\begin{figure}
\resizebox{\hsize}{!}{\includegraphics{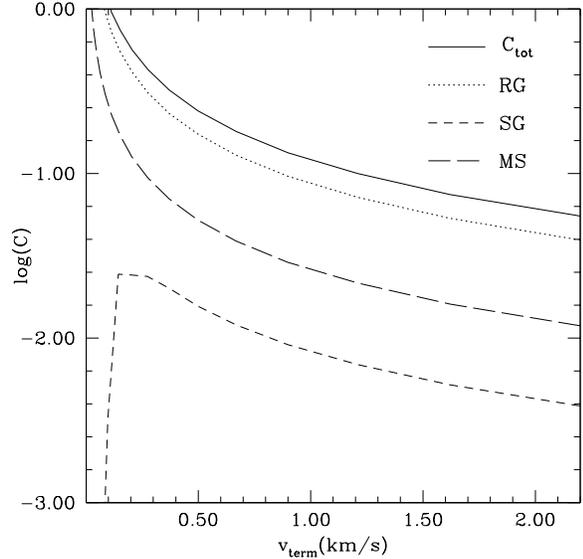}}
\caption{
The total value of the covering factor, $C_{\ma {tot}}$, defined as the sum of the 
contributions due to the envelopes of different stellar 
types, as well as the distinct contributions to this value from the various stellar types, 
are shown as  functions
of the terminal velocity  of expansion $v_{\ma{term}}$ (in km/s), characterizing the contributing envelopes. 
In the present ``intermediate'' case, both RGs and MSs give substantial contribution to the 
definition of the total covering factor}
\label{cvinterm}
\end{figure}

We want to note that, with respect to the case presented in the previous Section, 
for red giants the present choice
of mass loss rate function has significantly larger values over most of the region from which 
we expect to have BLR contributions. A similar condition holds for MS stars as well, but 
the effect is weaker than for RGs.
As for SG stars, the situation is less well defined, but $(\dot M_*)_{\ma{SG}}(r)$ 
is effectively larger than what we get in Sect.~5.2 over the outer portion of the 
region contributing to the BLR emission for SGs.
As a consequence of this, the resulting contributions of the different types of stars 
to the total covering factor, that we again show as functions of the expansion velocity 
parameter $v_{\ma{term}}$ in Fig.~\ref{cvinterm}, turn out to have a different relative importance with respect to 
the ``enhanced'' mass loss rate case of Sect.~5.2. 
In fact, in the present case RG stars are the dominant contributors to the total covering 
factor, but the contribution of main sequence stars is significant ($\sim 22\%$ of $C_{\ma{tot}}$)
 at any value of the 
velocity parameter $v_{\ma{term}}$.
On the contrary, the contribution due to  much less numerous SG
stellar component does not increase as much, and  we get a $C_{\ma{SG}}$ estimate which is 
$\leq 7\%$ of the total value of the covering factor for any $v_{\ma{term}}$.

\begin{figure}
\resizebox{\hsize}{!}{\includegraphics{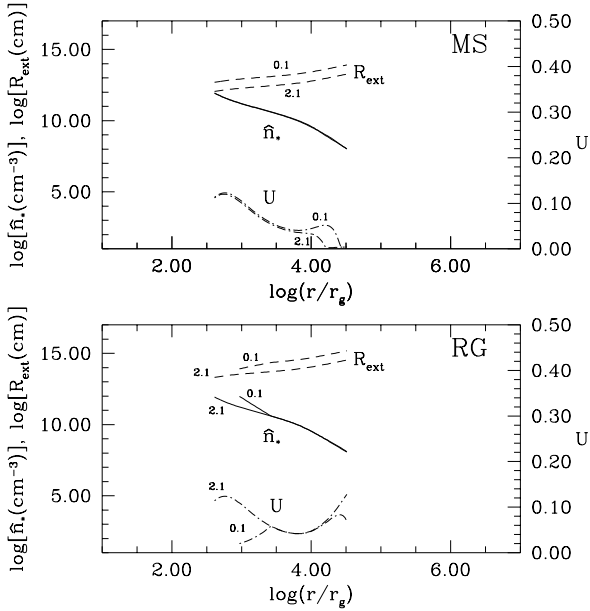}}
\caption{Physical parameters characterizing stellar envelopes contributing 
to the BLR as functions of the normalized distance $r/r_{\ma g}$ from the central black hole;
these are the mean density of the envelope, $\hat n_*(r)$, the estimate of the envelope
extension, $R_{\ma{ext}}(r)$, and the ionization parameter, $U(r)$.
The interval over which the physical quantities are plotted represents the ``BLR'' interval
$[r_1,r_2]$ as defined in Sect.~4.2; the labels 0.1 and 2.1 
identify the specific value (in km/s) of the terminal velocity parameter for which they are 
computed. 
We show the results for  both MSs (upper panel) and RGs (lower panel), since they 
both significantly contribute to the total covering factor in the present case of 
``intermediate'' mass loss rates.}
\label{intermfig}
\end{figure}

A closer inspection of the BLR models we have used to build the curves in Fig.~\ref{cvinterm} 
shows again that the bow-shock mechanism for confinement of the stellar expanding 
envelopes is the one that effectively determines the extension of the envelopes 
under any condition for MS stars. For RG envelopes this is still true, except 
for the very low velocity end of the interval shown, in which case other confinement
mechanisms dominate in the inner portion 
($r\simlt 500r_{\ma g}$) 
of the BLR contributing region, $[r_1,r_2]_{\ma{RG}}$. On the other hand, 
SG envelopes are confined basically by bow-shock mechanism for $v_{\ma{term}}> 0.3$~km/s, 
whereas for lower values of the velocity parameter other confinement mechanisms become more
efficient in defining the envelope extension, just like the case analysed in Sect.~5.2.

Regarding the definition of the range of values for the velocity parameter $(v_0)_{\ma i}$
for the present case, again we report representative values analogous to those 
defined in Sects.~5.1 and 5.2 for the previous mass loss rate cases, namely
$(v_0)_{\ma {MS}}(v_{\ma{term}}=0.1~{\rm km/s})= 5.8\times 10^{-4}$~km/s, 
$(v_0)_{\ma {RG}}(v_{\ma{term}}=0.1~{\rm km/s})= 3.7\times 10^{-4}$~km/s, 
$(v_0)_{\ma {SG}}(v_{\ma{term}}=0.1~{\rm km/s})= 8.8\times 10^{-4}$~km/s, and 
$(v_0)_{\ma {MS}}(v_{\ma{term}}=2.1~{\rm km/s})= 3.4\times 10^{-2}$~km/s, 
$(v_0)_{\ma {RG}}(v_{\ma{term}}=2.1~{\rm km/s})= 2.1\times 10^{-2}$~km/s, 
$(v_0)_{\ma {SG}}(v_{\ma{term}}=2.1~{\rm km/s})= 5.1\times 10^{-2}$~km/s.
We have given the values for main sequence stars as well, since 
they give a substantial contribution  to the covering factor 
also in this case.

Similarly to Figs.~\ref{standfig} and \ref{squarefig} for the previous two cases, 
Fig.~\ref{intermfig} shows the profiles 
of the relevant physical parameters for stellar envelopes contributing to the BLR as
functions of the distance $r$. We show the results pertaining to the same two 
representative values of  the velocity parameter $v_{\ma{term}}$, namely 
0.1~km/s and 2.1~km/s.
The extension of the BLR contributing interval for MS envelopes is 
$[r_1,r_2]_{\ma{MS}}=[410r_{\ma g},3.2\times 10^4r_{\ma g}]\simeq[6.8\times10^{15}$~cm,
$5.3\times10^{17}$~cm], independently of the velocity parameter, again because of the
fact that the extension of the envelopes is always defined by the bow-shock mechanism.
For RG envelopes, the BLR contributing interval has the same upper limit $r_2$ as that 
for MS contributors, whereas the inner boundary $r_1$ shows a dependence on the velocity
parameter (see previous paragraph).

As it is easy to see from Fig.~\ref{cvinterm}, the present choice of mass loss rates for the central cluster stars
turns out to be the most favourable one from the point of view of the value of the  covering factor 
that can be recovered with our models. In particular, the interval  of observationally inferred 
values for a Seyfert~1 BLR, defined in Sect.~4.2 to be ranging from 0.05 up to around 0.25,
can be obtained with stellar envelopes  characterized by expansion velocities in the range
$0.5-3.0$~km/s, which corresponds to subsonic, but physically sensible values for an 
expanding stellar atmosphere.
As for the values of $C_{\ma{forb}}$, they turn out to be sufficiently small to be
considered ``vanishing'', similar to what happens in the case for ``enhanced'' mass
loss rates of Sect.~5.2, since they at most reach $6\%$ of $C_{\ma{tot}}$.

As  a final consideration for the intermediate mass loss scenario
we estimate total number of emitting units. 
 Indeed, similar to the case for ``enhanced'' mass loss rates, we 
can include MS star envelopes in the total number of individual contributors to the 
model BLR, since their contribution to the total covering factor is also substantial in this
case. This allows an estimate of $(N_*)_{\ma {tot}}$ 
$\sim 3.2\times 10^7$ stars; this number is even a little higher that the one estimated for the 
case of ``enhanced'' mass loss rates; this is due to the fact that in the present
case the external boundary of the BLR contributing interval of distances is larger 
than that of the case discussed in Sect.~5.2, thus allowing to include a larger number 
of stars in the computation.

\medskip
\medskip

\section{General Discussion}

The  results presented in the previous section show that  our model is capable
of reproducing the observational requirements listed in Sect.~4.2  (points 1-4).
In fact,  from Fig.s \ref{standfig}, \ref{squarefig} and 
\ref{intermfig}  it is evident that the ionization parameter $U$ 
is in the expected range of values, while $r_{\ma{BRL}}$ is contained in the 
intervals [$r_1, r_2$] where the conditions a)-d)
of Sect.~4.2 are satisfied. As far as the covering factor is concerned, 
it is evident that its value strongly depends on the choice adopted for the
stellar mass loss rates (compare Fig. \ref{cvstan}, \ref{cvsquare}
 and \ref{cvinterm}). In the  case of enhanced and intermediate mass loss
rates, values in the
range 0.05- 0.25 can be recovered by choosing suitable values 
 of the stellar envelope expansion velocities, while for standard mass loss
rates this is possible only for very low velocity values. Hence, our values of $C_{\ma {tot}}$
tend to exclude stellar envelopes with standard mass loss rates as
plausible emitting units for a BLR model.
On the other hand,  regarding the  
covering factor for ``broad'' forbidden line emission, we recall
that the 
star density distribution we have adopted has been chosen so that in all the
presented cases  $C_{\ma {forb}}$ is minimized (see Sect.~5); in fact, it turns out 
to be negligible
in both the cases presented in Sects.~5.2 and 5.3. We recall here that 
in the ``standard'' mass loss rate case, $C_{\ma{forb}}$ is not negligible and we 
consider this one of the
reasons why that particular model is not viable for building up a realistic BLR model.

While the analysis of this paper  has produced  strong constraints 
to most of the free  model parameters,
for some, namely the stellar distribution, the mass loss rates and the expansion velocity, 
the model cannot  operate a definite choice.  Any further test of the model and
selection of these parameters requires more 
information coming
from the comparison between the observed line profiles and relative intensities, and 
those predicted by the model.
As we have already mentioned,
we have planned to compute the detailed line
profiles in a forthcoming paper, having now obtained a deeper insight in the role 
that different parameters
play in the model.  However, at this stage we would like to compare our
model with the results of some other authors (Goad \& Koratkar \cite{goad},
Kaspi \& Netzer \cite{kaspi}, Korista \& Goad \cite{korista00}) who made attempts to infer from line
profiles and intensities some ``average'' physical conditions in BLR condensations.

In the last years much work has been devoted to infer the BLR physical
configuration starting from the observed BLR properties. Since the BLR emitting gas has been shown
to be
stratified in density (see Peterson \cite{peterson93} and Netzer \& Peterson
\cite{netzer} for reviews)  the idea was to derive both the
dependence of  the characteristic
number density of the gas in the ``cloud-like'' emitting clumps ($\hat n_*(r)$
in our notation)  and
that  of the ``cloud-like'' structures' distribution ($\rho _*(r)$ in our notation)
on  the radial distance $r$ from the central black hole. This analysis has been applied
by the authors mentioned above
to the case of the well studied and representative
Seyfert 1 NGC 5548. 

Combining a photoionization code with an
optimization routine, 
Goad \& Koratkar (\cite{goad})  
have found  that line ratios and variability time scales can
be reproduced assuming $\hat n_*(r) \propto r^{-2}$. 
Kaspi \& Netzer (\cite{kaspi}), through the ``direct'' method of guessing
cloud properties and distribution and calculating the resulting
emission lines, have also  found that  a density gradient is necessary but their
best  fit is $\hat n_*(r) \propto r^{-\rm s}$ with $1\leq {\rm s} <1.5$.
In our results $\hat n_*(r)$  slope is variable in the interval [$r_1, r_2$]
where star envelopes contribute to build up the covering factor. For $\hat n_*(r)$ 
profiles shown in   
Fig.s \ref{standfig}, \ref{squarefig} and \ref{intermfig}  
a piece-wise power law approximation gives logarithmic slopes from 1.5 up to 2.
These values 
cannot be used to exclude or to prefer one of the assumed mass loss profiles,
since they do not contradict any of the two previously quoted analyses.

Recently Korista \& Goad (\cite{korista00}) have tested the ``locally optimally emitting
clouds'' (LOC) model, proposed by Baldwin et al. (\cite {bal95}), by comparing  the predicted 
spectrum with that of NGC 5548. One of the results they attain is that
the power law index of the quantity they define as the
 radial cloud covering fraction must be in the range -1.6 to -0.5.
In our notation this implies that the product  $R_{\ma {ext}}^2(r) \rho_*(r)$
must  be a function of $r$ which is representable by a power law whose exponent should be
in the range $[-1.6, -0.5]$ mentioned above. 
Since the star distribution function $\rho_*(r)$ that we have adopted has a logarithmic 
slope 
ranging from -0.7 around $r/r_{\ma g} =10^3$ down to -2 around $r/r_{\ma g} =10^4$, 
this comparison favours $R_{\ma {ext}}(r)$ functions flatter than $r^{0.75}$.
This fact again implies that ``enhanced'' and ``intermediate'' mass loss rates are preferable,
since in these cases the functional form of $R_{\ma {ext}}(r)$ does indeed fulfill the
requirement above (in fact, it turns out that it is $\propto r^{0.66}$ or flatter  
both for MS and RG stars).

Both Korista and Goad (\cite{korista00}) and Kaspi and Netzer (\cite{kaspi})  derive 
an estimate for the extension of the BLR.
For the first ones the maximum BLR extension can be up to 200~light days, while for the second
ones it can be up to 100~light days. In our model, the values for the derived 
external BLR radius, $r_2$, for the case of ``intermediate'' mass loss (see Sect.~5.3)
are larger than those derived for the case of ``enhanced'' mass loss rates presented in
Sect.~5.2; however, they are in any case consistent with the estimate obtained 
by Korista and Goad (\cite{korista00}). Again, this comparison is encouraging for our model, but
not definite enough to allow for a selection of most appropriate mass loss rate values.
In any case, the general comparison of our model with the global inferred physical 
conditions for the BLR discussed above  favours 
the two configurations analyzed in Sects.~5.2 and
5.3, characterized by distance dependent stellar mass loss rates and especially by 
a substantial contribution to BLR emission coming
from MS star envelopes, whose mass loss rate is strongly enhanced with respect 
to their ``standard'' value. 

Besides the X-ray illumination mechanism that we have discussed and examined in Sect.~5.2, 
other 
mechanisms have been analyzed in literature, that support the physical plausibility of this type 
of induced and $r$-dependent stellar mass loss  rates. Here we just want to mention 
a couple of these works. 
For example, MacDonald, Stanev \& Biermann (\cite{mcdonald}) performed an analysis of the effects 
of neutrino and high energy particle flux from the central source on the stars of the central
cluster and their winds; their main result is that these stellar winds can be affected in the 
sense of a mass loss enhancement. Also, Baldwin et al. (\cite{bal96}) have computed the radiative
acceleration on the  photospheric layers of stars at BLR distances for a number of high-luminosity
AGNs; comparing this acceleration with the star surface gravity, they conclude that 
all the stars, including main sequence ones, can indeed be affected, thus producing 
an enhanced mass loss.
We recall that for the case of $r$-dependent and enhanced mass loss rates that we have 
discussed in Sect.~5.3, 
we have not explicitly defined the physical mechanism inducing the enhancement, similarly to
what is done in Scoville and Norman (\cite{scoville}) (see Sect.~5.3), but we just  supposed 
one of such mechanisms to be at work.

Going back to what we have pointed out above, the two cases discussed in Sects.~5.2 and 5.3 fall 
in the same ``category'' of models, characterized by strongly enhanced mass loss rates and 
a significant contribution to the BLR emission coming from MS stars.
This is interesting from two different, but connected, points of view.
First, within this picture the large number of individual emitting units does not 
imply too large collision rates between the stars. In fact,  
the star collision rate (Eq. (\ref{tau})) relative  to the star distribution, $\rho_*(r)$, 
for which we have computed the models whose results are shown in Sect.~5 (the one 
labeled with ``1'' in Fig.~\ref{star}) turns out to be $\tau \simeq 1$~collisions/yr. 
The 
total star collision rate 
is an estimate of 
the total number of 
stellar collisions 
that take place 
in the whole cluster in a year;
however, to understand to what degree the cluster structure is affected
by stellar collisions another quantity must be introduced, that is
the time that it takes for a star
to be destroyed by collisions with other stars belonging to the cluster.
Following Begelman \& Sikora (\cite{begelman92}),  we introduce here the destruction
time 
due to mutual collisions 
for a star in the cluster 
as
\beq
t_{\ma {coll}}(r) \simeq {10 \over \rho_*(r) f_i~\pi (R_*)_i^2 V(r)},$$
\label{tcoll}
\eneq
where the number 10 at numerator is the number of collisions 
that a star has to undergo to be finally disrupted; the value used above 
is estimated from the fraction of 
stellar mass lost per collision that Begelman \& Sikora (\cite{begelman92}) 
evaluate as $\sim 0.1$. 
In our model 
the destruction time per star 
turns out to be the same for red giant and main sequence stars
(since $(R_*)_{\ma{RG}}=10~(R_*)_{\ma {MS}}$ and $f_{\ma {RG}}=0.01~f_{\ma {MS}}$) and 
it varies with the distance $r$ from the central black hole. To derive an estimate of its value, we have computed it
in the region of interest for this work, that is
at a distance which is around a half of the estimated external radius ($r_2$)  of the BLR model,
i.e. $\sim 10^4r_{\ma g}$,
 obtaining $t_{\ma {coll}}(x=10^4)\simeq 8\times 10^8 $yr.
A necessary condition for our picture to be consistent is that the MS star evolution time 
is shorter than the star destruction time. Taking into account the value of  $t_{\ma {coll}}$
derived above, this condition is verified  for stars with masses larger than $2.6~M_{\odot}$
which, in a star cluster of $3\times 10^7$ stars, assuming the Salpeter initial mass 
function, constitute  $\sim 3 \%$ of the total star 
number.  This percentage of stars can account for the assumed quantity
of evolved stars in our model. Obviously, going closer 
to the cluster center,
$t_{\ma {coll}}$ decreases  and this condition can no longer be fulfilled. 
However, 
there are a few arguments  that suggest 
that the destruction time for collisions computed above
should be considered as a sort of lower limit evaluation.
In fact, as Scoville \& Norman ({\cite{scoville}) argue, ordered stellar motions may result in a lower collision
rate.  In addition, as discussed by these same authors,  stars orbiting on elongated elliptical orbits  
would spend only a short part of
their life in the inner region of the cluster and this would result in a larger survival probability as well.
Our present discussion is centered on survival conditions for the central stellar cluster as we have chosen to 
model it. Its relation to the AGN lifetime (possibly $< 10^8$yr) and to the time at which the nuclear activity 
switches on with respect to the evolutionary stage of the cluster itself would indeed deserve further insight. However, 
in the present context, it is probably just the case to mention the possible relevance of this issue and we leave 
its analysis to different work. Back to the original point, 
even accounting for these last considerations, the discussion above 
shows that 
the very large values of the star number density characterizing the stellar distribution $\rho_*(r)$ that 
we have chosen for our models (the one labelled 
with 1 in Fig.(\ref{star}))
can be regarded as very close to the upper limit for the maintainance of the cluster integrity.
Indeed, higher stellar densities would result in a cluster evolution much faster than what is required 
when the cluster star envelopes are believed to represent the structural components of an AGN Broad Line Region.
As a matter of fact, this is indeed the case when the only stellar populations forming expanding envelopes suitable for
Broad Line emission (i.e., contributing to the BLR) are those of RGs and SGs. 
In fact, both RG and SG populations amount to only a small fraction of the total stellar number in the cluster (i.e., $
\sim 1\%$, see Sect.~4.1);
as a consequence, requiring a number density of RG and SG envelopes, as contributing emitting units,
sufficiently large to justify the observationally inferred covering factors implies 
a total number density of stars around 100 times larger. This would lead to a destruction time 
(as estimated through Eq.~(\ref{tcoll})) unacceptably  short.
(see Begelman \& Sikora \cite{begelman92}).

In addition, a second point of interest regarding the significant contribution of MS star envelopes to the BLR emission,
is that, owing to the large number of emitting
structures,  our model 
can be 
easily set in accordance with the
results of the studies of  
bright AGN spectra, performed with cross-correlation techniques, such as those of 
Arav et al. (\cite{{arav97},{arav98}}),
that  claim to determine a lower limit on the number of  individual emitting structures
of the BLR, for models  based on discrete emitting units composing the region. 
In particular, for Mk~335, these same authors derive a  lower limit
for the number of reprocessing cloud-like structures,
which is expected to be around $3\times 10^6$.

To compare our model to the results inferred by Arav et al.  (\cite{{arav97},{arav98}})
for Mk 335, we have to take into account configurations with
a more powerful central AGN nucleus. Therefore,  
we have analyzed models for $L= 10^{45}$~erg/s, that is the
luminosity of this source. To model an AGN with this luminosity,
we have chosen the same central black hole mass as that shown in Table~1 of Paper~I, namely 
$M_{\ma{BH}}= 1.12\times 10^8 ~M_{\odot}$; also, for simplicity, we suppose that the 
central star cluster has just the same profile $\rho_*(r)$
as the one adopted for the case of luminosity $L=10^{44}$~erg/s (see the curve labeled with
``1'' in Fig.~\ref{star}). 
With these hypotheses, we can test the resulting total number of stars within 
a distance corresponding to the estimated external boundary of the BLR model, 
comparing it with Arav et al. (\cite{{arav97},{arav98}})
lower limit evaluation of the number of individually emitting units in the source mentioned.
Our result turns out to be $(N_{\ma {*tot}})_{45}\simeq 3.9\times10^7$~stars, where we have 
computed this number using the estimate we obtain for 
$(r_2)_{\ma {MS}}\sim 2.3\times 10^4 r_{\ma g}$, since the total number of stars is essentially
determined by the number of MS stars  in the cluster for scenarii of the type
described in Sects.~5.2 and 5.3. Our evaluation of $(N_{\ma{*tot}})_{45}$
turns out to be well above the 
lower limit found by Arav et al. (\cite{{arav97},{arav98}}), and this is another encouraging
outcome.
As for other properties of broad line emitting stellar envelopes in this same case of 
``high'' luminosity ($L=10^{45}$~erg/s) AGN, these turn out to show behaviours that are 
similar to those we have discussed in Sects.~5.2 and 5.3 for the case for $L=10^{44}$~erg/s.
We note in passing that the bow shock mechanism for envelope confinement is in general still 
the most efficient, 
although in this higher luminosity case Comptonization confinement (see Sect.~2, Eq.~(\ref{rcomp})) can be 
dominant in the inner ($r\simlt 500r_{\ma g}$) 
portion of the model broad line emitting zone (that is, closer to the
central luminosity source, where the radiation flux is stronger). 

As a final remark, we note that  
the fact that the model outcome in this scenario  can be set in accordance with the 
very large number of discrete broad line emitters in the region
required by the analysis performed by Arav et al. (\cite{{arav97},{arav98}}) 
could be of relevance with respect to the BLR structure problem, because
it would show that discrete emitting units models of the BLR are not necessarily ruled out 
by the analysis of Arav et al. (\cite{arav98}). 

\medskip
\medskip

\section {Conclusions}

In this paper we have analyzed the scenario in which BLR  emission
 originates in expanding atmospheres of the stars of a dense central cluster
embedded in  the hot, tenuous  plasma
outflowing from the AGN.
Immersing the  central star cluster in its specific AGN environment and
investigating the interaction of the stellar wind with the AGN outflow,
we consider
the shock fronts generated by the interaction of the stellar wind and AGN
outflow 
as another confining mechanism for the stellar envelopes. In our model,
the nuclear wind structure depends on the 
features of the dense stellar cluster around the central black hole (including the
mass loss rates of the stellar components of the cluster itself). At the same time,
the physical conditions of the stellar envelopes emitting the broad lines depend
on the nuclear wind properties. Thus, looking for the stellar envelope physical conditions
that are appropriate to reproduce the general characteristics of the BLR as
they are deduced from observations, we have solved the problem consistently.

We have identified  those
parameters that significantly influence the 
resulting BLR model properties: these turn out to be $v_0$, a stellar wind velocity parameter, 
the set $\{\dot M_{\ma{MS}},\dot M_{\ma{RG}},\dot M_{\ma{SG}}\}$ of stellar mass loss rates, 
and $\rho_*(r)$, the star number density distribution.
Indeed, we have explored different possibilities for building a BLR composed of
mass losing stellar envelopes, determining, in the very specific and extreme
AGN central region  environment, the conditions under which different star types
contribute to the observed line emission. In summary, 
we have selected two main choices for the BLR structure, namely  1)~ a scenario in which
stars are taken as characterized by their ``standard''
({\it i.e.}, as known from solar neighbourhood
studies) mass loss rate, examined in Sect.~5.1, 
and  2)~ another one in which stars are  characterized by an
enhanced mass loss rate. Both cases, the ``enhanced'' mass loss rates and the
``intermediate'' mass loss rates, examined in Sects.~5.2 and 
5.3, belong to this second picture.
In the first scenario, red giant  and supergiant envelopes  are the foremost  contributors
to BLR, while in the second, main sequence stars
turn out to be an important constituent of the BLR.

For both scenarios, we have built configurations that meet
the basic physical conditions 
 for stellar extended atmospheres to be able to contribute to BLR emission.
As  discussed in Sect.~6, the observational requirements 
 are much better fulfilled by models built in the second scenario, allowing for 
large mass loss rates, with respect to those that we can construct starting from
the assumption of standard stellar mass loss rates. 
In addition, 
 if we compare the two scenarios with the further conditions
inferred from a global analysis of BLR observations 
 and with the deductions of Arav et al.
(\cite{{arav97},{arav98}}) on the total number of individually emitting structures in the BLR,
again we find a differentiation in behaviour and outcomes.
The scenario in which main sequence  stars are  also endowed with winds capable of
producing stellar envelopes with a non-negligible mass loss rate  is therefore preferable.
This point shows the importance  of the choice of the mass loss rate in our model.

 Besides the above general considerations, 
in our model the most stringent quantity turns out to be $C_{\ma{tot}}$. In particular,
covering factors that match the range derived from
observations are not easily obtained.
To recover values of $C_{\ma {tot}}$ as large as those observationally inferred 
we need to suitably choose the star number density $\rho_*(r)$
and/or the ratio $\dot M/v_0$.
The star number density distribution of the
cluster orbiting around the central black hole turns out to be crucial in our model.
 In fact, 
the observed BLR features can not  be recovered for any given $\rho_*(r)$: 
 while the required global amount of stars in the cluster is not  large,
 the star number density in the central region (i.e. around $10^{16}$ cm) must be rather high.
Also, high mass loss rates and low outflow velocities are preferable. 

Even with these limitations, the model nevertheless 
constitutes a consistent structure for a BLR embedded in an AGN  wind-type plasma outflow,
with individual emitting units characterized by the appropriate values of physical quantities
for broad line emission.
Of course, a more detailed comparison of the line shape and intensity produced by our model
would be necessary to strengthen the validity of this approach.
As a first step, however, we can assert that
the results we obtain  are encouraging, because this analysis has determined  strong
constraints on the free parameters in the model. 
 Such constraints will be useful in the derivation of the characteristic line profiles that 
can be predicted by our model.

\medskip
\medskip

\begin{acknowledgements}
We thank Nick Scoville and Leslie Hunt for helpful discussions; it is also
a pleasure to acknowledge Leslie Hunt's  careful reading of the paper. This work was 
partly supported by the Italian Ministry for University
and Research (MURST) under grant Cofin00-02-36.
\end{acknowledgements}


\begin{thebibliography}{}

\bibitem[1994]{alexander94}
Alexander, T., Netzer, H., 1994, MNRAS 270, 781 (AN1)

\bibitem[1997]{alexander97}
Alexander, T., Netzer, H., 1997, MNRAS 284, 967 (AN2)

\bibitem[1997]{alexander97b}
Alexander, T., 1997, MNRAS 285, 891

\bibitem[1985]{allen}
Allen, C.W., 1985, Astrophysical Quantities, The Athlone Press, London

\bibitem[1997]{arav97}
Arav, N. et al. 1997, MNRAS 288, 1015

\bibitem[1998]{arav98}
Arav, N. et al. 1998, MNRAS 298, 990


\bibitem[1995]{bal95}
Baldwin, J. et al., 1995, ApJ 455, L119

\bibitem[1996]{bal96}
Baldwin, J. et al., 1996, ApJ 461, 664 

\bibitem[1997]{Baldwin97}
Baldwin, J.A., 1997, in: Emission Lines in Active Galaxies: New Methods and Techniques,
              ASP Conference Series, Vol.113,
            eds.  B.M. Peterson, F.Z. Cheng and A.S. Wilson (San Francisco:ASP), pag.80

\bibitem[1992]{begelman92}
Begelman, M.C., Sikora, M., 1992, in ``Testing the AGN Paradigm'', Proc. AIP Conf. 254, 
Holt, S.S., Neff, S.G., Urry, C. M. eds., AIP, New York, p. 568

\bibitem[1999]{crenshaw99b}
Crenshaw, D.M., Kraemer, S.B.,Boggess, A. et al., 1999, ApJ 516, 750

\bibitem[1987a]{david87a}
David, L.P., Durisen, R.H., Cohn, H.N., 1987a, ApJ 313, 556

\bibitem[1987b]{david87b}
David, L.P., Durisen, R.H., Cohn, H.N., 1987b, ApJ 316, 505

\bibitem[1994]{elvis94}
Elvis, M. et al., 1994, ApJS 95, 1

\bibitem[1998]{goad}
Goad, M.R., Koratkar, A., 1998, ApJ 495, 718

\bibitem[1993]{hameury93}
Hameury, J.M., King, A.R., Lasota, J.P., Raison, F., 1993, A\&A, 277, 81

\bibitem[1997]{kaspi97}
Kaspi, S., 1997, in : Emission Lines in Active Galaxies: New Methods and Techniques,
       ASP Conf. Ser. vol. 113,
		eds. B.M. Peterson, F.-Z. Cheng,\& A.S. Wilson, (San Francisco:ASP), 159

\bibitem[1999]{kaspi}
Kaspi, S., Netzer, H., 1999, ApJ 524, 71

\bibitem[2000]{kaspi00}
Kaspi, S. et al., 2000, ApJ 533, 631

\bibitem[2001]{kaspi01}
Kaspi, S. et al., 2001, ApJ 554, 216

\bibitem[1989]{kazanas}
Kazanas, D., 1989, ApJ 347, 74


\bibitem[1999]{korista}
Korista, K., 1999, in: Quasars and Cosmology, ASP Conference Series, eds.
G. Ferland, J. Baldwin (San Francisco:ASP) (astro-ph/9812043)

\bibitem[2000]{korista00}
Korista, K., Goad, M.R., 2000, ApJ 536, 284

\bibitem[1981]{krolik}
Krolik, J.H., McKee, C.F., Tarter, C.B., 1981, ApJ 249, 422

\bibitem[1999]{krolik99}
Krolik, J.H., 1999, Active Galactic Nuclei, Princeton University Press

\bibitem[2001]{krolik01}
Krolik, J.H., Kriss, G.A., 2001, ApJ 561, 684

\bibitem[2002]{krolik02}
Krolik, J.H., 2002, astro-ph/0204418

\bibitem[1999]{lamers}
Lamers, H.J.G.L.M., Cassinelli, J.P., 1999, Introduction to Stellar Winds, Cambridge University
Press, Cambridge

\bibitem[1991]{mcdonald}
MacDonald, J., Stanev, T., Biermann, P.L., 1991, ApJ 378, 30 

\bibitem[1991]{murphy}
Murphy, B.W., Cohn, H.N., Durisen, R.H., 1991, ApJ 370, 60 

\bibitem [1990]{netzer90}
Netzer, H., 1990, in: Active Galactic Nuclei, 
   T.,J.-L., Courvoisier, M., Mayor eds., Springer Verlag, Berlin, p. 57

\bibitem [1997]{netzer}
Netzer, H., Peterson, B.M., 1997, in Astronomical Time Series, Proceedings of the Wise
Observatory 25$^{th}$ Ann. Symp., eds. D. Maoz,
A. Sternberg, E. Leibovitz, Kluwer Academic Publishers, Dordrecht
(astro-ph/9706039)

\bibitem[1985]{perry}
Perry, J.J., Dyson, J.E., 1985, MNRAS 213, 665

\bibitem[1993]{peterson93}
Peterson, B.M., 1993, PASP 105, 247

\bibitem[1997]{peterson97}
Peterson, B.M., 1997, ``Introduction to Active Galactic Nuclei'', Cambridge University
Press, Cambridge


\bibitem [2000]{pietrini}
Pietrini, P., Torricelli-Ciamponi, G.,  2000, A\&A 363, 455 (Paper I)

\bibitem[1989]{rees89}
Rees, M.J., Netzer, H., Ferland, G.J., 1989, ApJ 347, 640

\bibitem [1988]{scoville}
Scoville, N., Norman, C., 1988, ApJ 332, 163 


\bibitem [1999]{taylor}  
Taylor, J.A., 1999, Ph.D. Thesis ``Line Emission from Stellar Winds in AGNs'',
  University of Maryland, College Park


\bibitem[1990]{buren}
Van Buren, D. et al., 1990, ApJ 353, 570

\bibitem[1999]{wandel99}
Wandel A., Peterson, B.M., Malkan, M.A., 1999, ApJ 526, 579

\bibitem [1997]{weymann97}
Weymann, R.J., Morris, S.L., Gray, M.E. et al., 1997, ApJ 483, 717


\end{thebibliography}
\end{document}